\documentclass[journal, comsoc]{IEEEtran}
\usepackage{graphicx}
\usepackage{color}
\usepackage{cite}
\usepackage{algorithm}
\usepackage{algorithmicx}
\usepackage{algpseudocode}

\usepackage{url}
\usepackage{subfigure}
\usepackage{bm}
\usepackage{optidef}
\usepackage{amsmath}
\usepackage{amssymb}
\usepackage{amsfonts}

\graphicspath{{./fig/}}

\renewcommand{\algorithmicrequire}{\textbf{Input:}}
\renewcommand{\algorithmicensure}{\textbf{Output:}}

\newcommand{\aircomp}[1]{\mathrm{AirComp}\left({#1}\right)}

\begin{document}
\title{Adaptively Weighted Averaging Over-the-Air Computation and Its Application to\\Distributed Gaussian Process Regression}

\author{Koya~Sato, \IEEEmembership{Member,~IEEE} and Koji Ishibashi,~\IEEEmembership{Senior Member,~IEEE}% <-this % stops a space
        \thanks{
        A part of this work was presented at the IEEE GLOBECOM Workshops\cite{ksato-gcw2022}.
        This was supported in part by JST CRONOS Grant Number JPMJCS24N1, JST AdCORP Grant Number JPMJKB2307, and JSPS KAKENHI Grant Number 22K14255.\\
        K. Sato is with the Artificial Intelligence eXploration Research Center, The University of Electro-Communications, 182-8585, Tokyo, Japan (e-mail: k\_sato@ieee.org).\\
        K. Ishibashi is with the Advanced Wireless and Communication Research Center (AWCC), The University of Electro-Communications, Tokyo 182-8585, Japan (e-mail: koji@ieee.org)\\
        Corresponding author: \it{Koya Sato}
        }
        }% <-this % stops a space
        % \markboth{Under Review}%
        \markboth{ACCEPTED FOR PUBLICATION IN IEEE Transactions on Cognitive Communications and Networking (DOI: 10.1109/TCCN.2025.3562345)}
        {Shell \MakeLowercase{\textit{et al.}}: Bare Demo of IEEEtran.cls for Journals}
%make the title area
\maketitle
\begin{abstract}
  This paper introduces a noise-tolerant computing method for over-the-air computation (AirComp) aimed at weighted averaging, which is critical in various Internet of Things (IoT) applications such as environmental monitoring. Traditional AirComp approaches, while efficient, suffer significantly in accuracy due to noise enhancement in the normalization by the sum of weights. Our proposed method allows nodes to adaptively truncate their weights based on the channel conditions, thereby enhancing noise tolerance. Applied to distributed Gaussian process regression (D-GPR), the method facilitates low-latency, low-complexity, and high-accuracy distributed regression across a range of signal-to-noise ratios (SNRs).
  We evaluate the performance of the proposed method in a radio map construction problem, which involves visualizing the radio environment based on limited sensing information and spatial interpolation.
  Numerical results show that our approach maintains computational accuracy in low-SNR scenarios and achieves performance close to ideal conditions in high SNR environments.
  In addition, a case study targeting a federated learning (FL) system demonstrates the potential of our proposed method in improving model aggregation accuracy, not only for D-GPR but also for FL systems.
\end{abstract}


\IEEEpeerreviewmaketitle

\section{Introduction}
\label{sect:intro}
\IEEEPARstart{E}{nvironmental} monitoring has been attracting attention in various fields, such as agriculture, disaster prevention, and radio map construction in wireless systems\cite{wang-iotj2024,rezaeibagha-iotj2021, bail-environhazards2021,cirillo-iotj2020,minelva-procieee2020,bi-wirelesscommun2019}.
In such a system, distributed Internet of Things (IoT) devices monitor the environment (e.g., temperature, humidity, and communication quality) and upload the data to a centralized server.
The server can then analyze real-time environmental information using statistical tools, including machine learning and spatio-temporal prediction models\cite{cressie2011}.
However, enhancing monitoring accuracy typically leads to a higher density of sensor deployments, which, in turn, causes several issues concerning communication and computation.
For example, collecting large-scale sensing data increases communication latency and consumes limited spectrum resources; further, processing large-scale data leads to a computational delay at the server.
Consequently, the design of efficient monitoring systems becomes critical, particularly for applications that demand both high accuracy and low latency.
\par
Recently, over-the-air computation (AirComp) has gained attention as a promising technique for the low-latency joint computation and communication of IoT systems\cite{sahin-comst2023,zhu-wc2021}.
AirComp leverages the superposition nature of analog signals. The transmitters modulate the message based on a pre-processing function, and then the base station (BS) decodes the computation result by a corresponding post-processing function. 
This post-processed sum of pre-processed messages is called \emph{nomographic function}, which enables efficient computation for distributed messages, such as summation or product, using only one slot, regardless of the number of nodes \cite{goldenbaum-twc2015}.
\par
There are, however, several drawbacks to AirComp-based systems regarding noise tolerance.
Consider an environmental monitoring scenario wherein multiple distributed nodes collect environmental data across various locations; e.g., radio map construction in wireless communication systems\cite{bi-wirelesscommun2019}. A radio map is a tool to visualize the communication quality of interest based on the sensing results and spatial interpolation, which can be optimized using Gaussian process regression (GPR).
Interpolation in regions lacking observations is a crucial task for the accurate analysis of environmental conditions over expansive areas.
Generally, such interpolation can be realized by weighted averaging of distributed data, focusing on the fact that the spatial correlation between samples depends on distance (e.g., inverse distance weighting (IDW)\cite{shepard1968two} and Kriging (or GPR)\cite{cressie-1990}).
AirComp can realize the weighted average through two communication slots and division of their communication results: (i) calculate the weighted sum of the sensing values, (ii) calculate the sum of weights, and (iii) normalize the weighted sum by the sum of weights.
Many regression analyses in environmental monitoring, not limited to the radio map construction problem, are designed based on such weighted averaging. Therefore, when aiming to improve the efficiency of data collection and analysis in environmental monitoring, designing the weighted averaging operation becomes even more crucial than merely implementing the summation operation.
\par
However, normalizing by the sum of weights often degrades the calculation accuracy due to the noise enhancement.
To illustrate this problem, consider a simple weighted averaging task involving two distributed nodes with positive weights $w_1$ and $w_2$, and real-valued measurement data $s_1$ and $s_2$, respectively.
The ideal weighted averaging operation can be expressed as $(w_1s_1 + w_2s_2) / (w_1 + w_2)$.  
When using AirComp, the sum of weights $(w_1 + w_2)$ and the weighted sum $(w_1s_1 + w_2s_2)$ must be computed over two slots based on independent AirComp operations, and the two results then need to be divided.  
Assuming that the computational error in the sum of weights is $\epsilon_w$ and the error in the weighted sum is $\epsilon_s$, the BS can estimate the computation result by $(w_1s_1 + w_2s_2 + \epsilon_s) / (w_1 + w_2 + \epsilon_w)$.  
Clearly, the noise term $\epsilon_w$ can severely degrade the accuracy of the weighted averaging computation due to the division operation.
\par
This issue can be avoided if the sum of weights is available at the BS before AirComp, as the noiseless sum of weights allows the computation $(w_1s_1 + w_2s_2 + \epsilon_s) / (w_1 + w_2)$. However, this approach is not practical in monitoring applications.
Let us consider a spatial monitoring application based on sensing by IoT devices and IDW-based spatial interpolation. In this scenario, the IoT devices upload sensing results to the server.
The server then interpolates the missing values by weighted averaging. To take spatial correlation into account, the IDW sets each weight factor as the inverse of distance from the interpolated location.
Thus, avoiding noise enhancement (i.e., computing the sum of weights at the BS alone) requires the BS to acquire precise information on all locations regarding sensing nodes.
    An alternative approach to mitigating this noise enhancement issue is to adopt simple averaging instead of weighted averaging; i.e., we can avoid effects of $\epsilon_w$ by $(s_1 + s_2 + \epsilon_s) / 2$, which can be realized by one AirComp operation for $s_1+s_2$.
    However, while simple averaging is noise-tolerant under low signal-to-noise ratio (SNR) conditions, its accuracy deteriorates compared to pure weighted averaging at high-SNR conditions.
    This example highlights the primary challenge in AirComp-based weighted averaging: \textit{how to mitigate the effects of noise enhancement while preserving the accuracy of the weighted averaging operation.}
\par
In this paper, we propose a noise-tolerant but highly accurate adaptive weighted averaging based on AirComp for low-latency monitoring systems.
In particular, the proposed method introduces an adaptive weighting function that truncates the maximum and minimum of the weights for the noise tolerance (i.e., reducing $\epsilon_w$ in the above example).
Pure AirComp determines a coefficient for the transmission power control so that the node with the worst channel condition transmits its message vector with the maximum transmission power.
When the channel condition is flat among each uplink, this coefficient tends to decrease the transmission power if the variance in norm values of message vectors is significant.
It reduces the received SNR in the sum of weight computation (i.e., enhancing $\epsilon_w$ in the above example), resulting in noise enhancement in the weighted averaging computation process.
% To mitigate this issue, our method introduces an adaptive weighting function that truncates the maximum and minimum of the weights based on the channel condition.
By optimizing the truncation parameters in this function via Bayesian optimization (BO)\cite{shahriari-ieee2016}, the proposed weighting function adapts the noise tolerance according to the SNR at the server: it works as an average function approximately in the low SNR region and as a pure weighted averaging function in the high SNR region.
\par
To demonstrate how the proposed method improve a practical computation problem, the proposed method is applied to distributed Gaussian process regression (D-GPR)\cite{pmlr-v37-deisenroth15}, which is a non-parametric distributed regression method.
The performance is evaluated in a radio map construction scenario, a typical application of GPR in wireless communication systems\cite{bi-wirelesscommun2019}.
An accurate radio map can provide the wireless system operator with precise statistical information about the wireless system, thereby enhancing overall system performance.  
Although many researchers consider it a promising technology for beyond 5G/6G systems, there is a trade-off between the amount of sensor data and performance improvement, motivating us to develop a more efficient radio map construction method.  
Our proposed AirComp-aided D-GPR enables the system operator to construct a radio map within limited communication slots, regardless of the amount of sensing data. Thus, this task highlights a potential application of the proposed method.
Our results show that the proposed method achieves low latency, low complexity, and accurate distributed regression.
\par
The major contributions of this paper are listed as follows.
\begin{itemize}
    \item We propose an adaptively weighted averaging method for AirComp-based distributed averaging. BO is applied to adjust the adaptive weighting function to mitigate the effects of noise on the estimation accuracy.
    \item The proposed method is applied to D-GPR\cite{pmlr-v37-deisenroth15} for low-latency, low-complexity, and accurate distributed regression. Numerical results in the radio map construction task demonstrate the root mean squared error (RMSE) of the proposed method is equivalent to the pure weighted averaging method in the high SNR region and approaches the simple averaging method in the low SNR region.
    \item In addition, we apply the proposed method to a federated learning (FL) system based on Federated Averaging (FedAvg)\cite{McMahan_AISTATS2017} to further explore applications of the proposed method. Numerical results show that the proposed method achieves stable and accurate training performance in both low and high SNR conditions.
\end{itemize}

\subsection{Related Works}
The central concept of the AirComp is based on the superposition property of electromagnetic field and nomographic function.
Nomographic function can be expressed as a post-processed sum of pre-processed sensor readings. By applying a non-linear function to the superimposed (and suitably pre-processed) signals, the receiver can obtain the desired function of the sensor readings with only one communication slot: e.g., arithmetic mean, weighted sum, and approximation of product\cite{sahin-comst2023}.
The origin of AirComp was introduced by \cite{goldenbaum-tsp2013}.
They discussed which functions are analog-computable at a single fusion center by harnessing the interference property of the wireless channel; further, the work in \cite{goldenbaum-twc2015} proposed reliable computation of nomographic function for clustered Gaussian sensor networks.
These works established the theoretical foundation of AirComp in the ideal wireless channels.
\par
Influenced by exponential growth in IoT systems, recent studies have discussed the practical implementation of AirComp in a wide range of viewpoints, such as extension to multiple-input multiple-output (MIMO)\cite{zhu-iotj2019,zhai-tcom2021}, reconfigurable intelligent surface (RIS)-aided design\cite{fang-tcom2021}, and robustness to imperfect (or statistical) channel state information (CSI)\cite{zhang-lcomm2022,jing-twc2022}.
Most of these studies are concerned with increasing the accuracy of nomographic function computation, designing communication parameters or modulation/demodulation functions with the goal of minimizing the mean squared error (MSE) of the computation results.
\par
From the application perspective, AirComp has been expected to be a promising component for FL systems in wireless channels.
FL is a distributed machine learning framework that enables multiple edge devices to collaboratively train a machine learning model without sharing raw data\cite{McMahan_AISTATS2017,nguyen-comst2021}.
By synthesizing distributed models on the server side, a large amount of data can be trained while ensuring data privacy.
There has been extensive research on applications of FL systems, including indoor positioning\cite{zuo_tmc2024} and vehicle-to-vehicle communication\cite{posner_ieeenw2021}, establishing it as a critical technology in next-generation mobile computing.
In FL, the training model is exchanged multiple times between edge devices and the server.
Therefore, communication delays during the training process pose a significant challenge, specifically in wireless channels, which has motivated discussions on communication design.
For example, FL can be accelerated by adaptively selecting training clients based on factors such as the communication channel conditions, computational capabilities, and data of each device\cite{fu_iotj2023, nishio_icc2019}.
Further, studies on integrated hardware and computing design have also been conducted; for instance, the authors in \cite{zuo_comml2024} introduce fluid antennas, demonstrating that optimizing antenna positioning and computational conditions jointly can significantly improve computational efficiency.
\par
However, these studies primarily assume digital communication systems.
When collecting updated models from multiple clients, the orthogonality of communication channels among clients is crucial, thereby degrading communication delays as the number of mobile devices increases.
In this context, model aggregation in FL, as exemplified by FedAvg\cite{McMahan_AISTATS2017}, is typically performed based on the averaging of local models.
Thus, it is expected that AirComp can reduce communication latency by jointly performing model aggregation and synthesis.
The work in \cite{yang-twc2020} proposed a joint device selection and beamforming design for AirComp-supported FL in wireless multiple-access channels.
Applying AirComp for privacy improvement in the FL was discussed in \cite{koda-globecom2020}\cite{park-twc2023}.
The result of the model aggregation by AirComp tends to be distorted by receiver noise and multipath fading. While this is not desirable in terms of training accuracy, it can suppress the amount of training data leaked from the model.
Data privacy in the FL can be thus guaranteed by adjusting the model's transmission method so that the amount of distortion in the model's synthesis results in achieving differential privacy\cite{dwork-tcs2014}.
\par
Some works discuss weighted aggregation for local training models in AirComp FL\cite{azimi-twc2024} or its decentralized setting\cite{zhai_twc2024,zhai_icc2024}. These works design the weights based on training conditions, including channel conditions\cite{azimi-twc2024} or the network topology\cite{zhai_twc2024,zhai_icc2024}.
While these strategies can enhance training performance, the sum of weights is pre-constrained to $1$, meaning it is known in advance on the server side. Consequently, they cannot address the noise enhancement problem, which motivates us to consider a more general setting; namely, weighted averaging, where the sum of weights is not predetermined.
% Although there has been a wide range of discussions concerning AirComp, to the best of our knowledge, this work is the first attempt to tackle the noise enhancement problem of weighted averaging in AirComp.
Weighted averaging is a widely used function, especially in monitoring applications; it will become more important in the IoT era. Through this work, we aim to expand the possible applications of AirComp.

\subsection{Organization}
The rest of this paper is organized as follows.
Sect.\,\ref{sect:systemmodel} describes the system model, including the computation target, channel model, and AirComp model.
Sect.\,\ref{sect:pure_weighted_averaging} introduces the AirComp-based pure weighted averaging method as a baseline.
After Sect.\,\ref{sect:proposed} presents the proposed adaptively weighted averaging method, the proposed method is applied to D-GPR in Sect.\,\ref{sect:gpr}.
Then, to explore further applications of the proposed method, Sect.\,\ref{sect:fl} presents a case study in the AirComp-based FL system.
Finally, Sect.\,\ref{sect:conclusion} concludes this paper.
\par
{\it Notations:} throughout this paper, the transpose, determinant, and inverse operators are denoted by $(\cdot)^\mathrm{T}, \mathrm{det}(\cdot)$ and $(\cdot)^{-1}$, while the expectation and the variance are expressed by $\mathbb{E}[\cdot]$ and $\mathrm{Var}[\cdot]$, respectively.
Further, $|\cdot|$ and $||\cdot||$ are defined as operators to obtain the absolute and Euclidean distance, respectively.
Finally, we define $\hat{(\cdot)}$ as the estimated result for the decorated variable.
\section{System Model}
\label{sect:systemmodel}
This paper investigates AirComp-based weighted averaging across distributed nodes.
This section begins by introducing the computation target.
Then, we introduce the signal model for AirComp and describe AirComp for the summation operation as a fundamental function in the weighted averaging.

\subsection{Computation Target}
Suppose that $M$ nodes are connected to a BS. The $i$-th node has the measurement vector ${\bf s}_i \in \mathbb{R}^{L}$ and the weighted vector ${\bf w}_i \in \mathbb{R}^{L}$ where $L$ is the message length.
Elements in ${\bf w}_i$ and ${\bf s}_i$ follow probability density functions (PDFs) $f_{w}(\cdot) (w \geq 0)$ and $f_{s}(\cdot)$, respectively.
\par
The task is to obtain an $L$-dimensional vector $\boldsymbol{\phi}$, whose $l$-th element represents the weighted averaging of the $l$-th samples.
This value is given by
\begin{equation}
    \phi_l = \frac{1}{\sum_{i=1}^{M}w_{il}} \sum_{i=1}^{M} w_{il} s_{il}.
    \label{eq:weighted_averaging}
\end{equation}
This weighted averaging often appears in sensing and estimation applications, such as IDW in spatial interpolation\cite{shepard1968two}, cooperative sensing in spectrum sharing\cite{akyildiz-phycom2011}, and distributed regression based on GPR\cite{pmlr-v37-deisenroth15}.

\subsection{Signal Transmission Model}
\label{subsec:signalmodel}
\begin{figure}[t]
    \centering
    \includegraphics[width=1.0\linewidth]{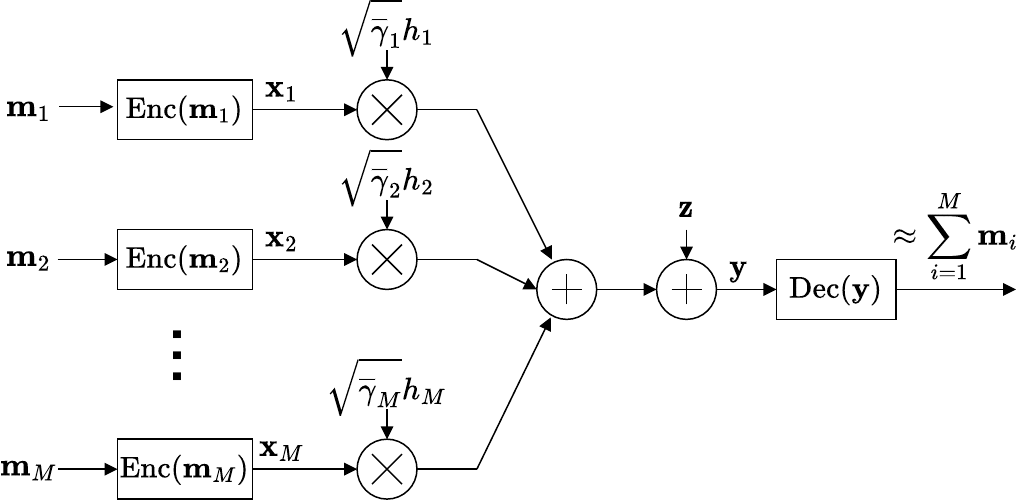}
    \caption{Signal transmission model for AirComp in a summation computation.}
    \label{fig:signalmodel}
\end{figure}
Fig.\,\ref{fig:signalmodel} summarizes the signal transmission model, where all nodes simultaneously transmit their messages to a BS over a shared wireless channel.
Let us consider a network with $M$ nodes synchronized in time, where the $i$-th node has a message vector $\mathbf{m}_i \in \mathbb{R}^L$.
Based on the AirComp, the BS tries to obtain the sum of these message vectors; i.e.,
\begin{equation}
    {\bf r} = \sum_{i=1}^{M}\mathbf{m}_i.
    \label{eq:sum_true}
\end{equation}
\par
The channel is modeled as either additive white Gaussian noise (AWGN) or independent and identically distributed (i.i.d.) Rayleigh fading, with the channel gain assumed constant during a single transmission.
Each node encodes its message $\mathbf{m}_i$ such that the BS can extract the sum of all messages; this operation is denoted by $\mathbf{x}_i = \mathrm{Enc}(\mathbf{m}_i)$.
The received signal at the BS is then given by
\begin{equation}
 \mathbf{y} = \sum_{i=1}^{M}\sqrt{\overline{\gamma_i}}h_i \mathbf{x}_i + \mathbf{z},
\end{equation}
where $\sqrt{\overline{\gamma}_i} \in \mathbb{R}$ is the average channel gain corresponding to large-scale fading such as path loss and shadowing, and $h_i$ is the instantaneous channel gain, which is assumed to be $1$ in the AWGN channel, while it follows $\mathcal{CN}(0, 1)$ in the Rayleigh fading channel.
Further, $\mathbf{x}_i$ is the transmitted vector constrained by the maximum transmission power $P_\mathrm{max}$ as $||\mathbf{x}_i||^2\leq P_\mathrm{max}$, and $\mathbf{z}\in \mathbb{C}^L$ is the AWGN vector following $\mathcal{CN}(0, \sigma_z^2)$, where $\sigma_z^2$ is the noise floor.
Then, BS extracts the sum of $\mathbf{m}_i$ using a decoding operation, defined by $\mathrm{Dec}(\mathbf{y})$.
\par
Note that this paper assumes that the BS and the nodes have perfect channel state information (CSI) about $\{\sqrt{\overline{\gamma}_i}h_i \mid i=1, 2, \cdots, M\}$ and the $i$-th node knows $\sqrt{\overline{\gamma}_i}h_i$ through the protocol detailed in the next subsection.

\subsection{AirComp for Summation Operation}
\label{subsec:aircomp_summation}
\begin{figure}
    \centering
    \includegraphics[width=1.0\linewidth]{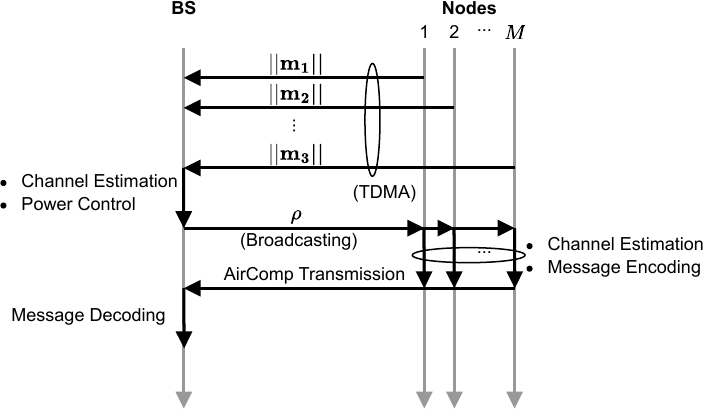}
    \caption{Timeline of AirComp for a summation computation.}
    \label{fig:timeline_aircomp}
\end{figure}

We next introduce the detailed design of $\mathrm{Enc}(\cdot)$ and $\mathrm{Dec}(\cdot)$ for the summation operation\cite{sahin-comst2023}, which is a fundamental operation for the weighted averaging\footnote{
    One might question the difference between AirComp and physical-layer network coding (PNC).
    PNC leverages waveform superposition to enable the simultaneous transmission of combined messages to be efficiently relayed over a wireless network, thereby enhancing the network throughput\cite{cheng_ieeenetwork2020}.
    However, in PNC-aided networks, the receiver must extract its intended information from the overlapped signals, which contrasts with the purpose of AirComp, which aims to estimate a function of the received signals\cite{wang_jiot2024}.}.
The brief timeline of AirComp-based summation is shown in Fig.\,\ref{fig:timeline_aircomp}.
Initially, each node transmits the quantized norm of its message $||\mathbf{m}_i||$ to the BS using digital modulation such as quadrature amplitude modulation in a time-division multiple access (TDMA) manner. Since the transmitted frame must include known pilot signals for synchronization, BS can obtain the instantaneous uplink CSI $\{\overline{\gamma}_i h_i \mid i=1, 2, \cdots, M\}$ while decoding $||\mathbf{m}_i||$, and then calculate a normalization coefficient $\rho$ for the power control. This factor is determined to ensure that the node with the smallest ratio of CSI to message norm, namely minimum of $\left(\sqrt{\overline{\gamma_i}}|h_i| / ||\mathbf{m}_i||\right)$, transmits the message at maximum transmit power, while preserving the relative order of magnitudes among the messages.
Thus, the BS calculates $\rho$ by
\begin{align}
    \sqrt{\rho} = \min \left\{\frac{\sqrt{\overline{\gamma}_i}|h_i|\sqrt{P_\mathrm{max}}}{||\mathbf{m}_{i}||} \;\middle|\; i=1, 2, \cdots, M\right\}.
    \label{eq:powercontrol_perfect}
\end{align}
BS next broadcasts $\rho$ to the nodes, and the $i$-th node estimates $\overline{\gamma}_i h_i$ while decoding $\rho$.
\par
For the AirComp transmission, the $i$-th node then converts its message $\mathbf{m}_{i}$ to a complex signal based on \emph{channel inversion principle}, so that
\begin{align}
    \mathbf{x}_{i} = \mathrm{Enc}\left(\mathbf{m}_{i}\right) \triangleq\frac{\sqrt{\rho}}{\sqrt{\overline{\gamma}_i}h_i} \mathbf{m}_{i}.
    \label{eq:encode_perfect}
\end{align}
\par
After the nodes transmit their messages simultaneously, BS receives the aggregated signal, which is derived by
\begin{equation}
  \mathbf{y} = \sqrt{\rho}\sum_{i=1}^{M} \mathbf{m}_i + \mathbf{z}.
  \label{eq:received_signal_sum}
\end{equation}
%% decoding

Next, we define the decoding function $\mathrm{Dec}(\cdot)$. Specifically, the BS attempts to extract the sum of message vectors from the received vector $\mathbf{y}$ by applying
\begin{equation}
    \mathrm{Dec}\left(\mathbf{y}\right) \triangleq \mathrm{Re}\left(\frac{\mathbf{y}}{\sqrt{\rho}}\right),
\end{equation}
where $\mathrm{Re}(\cdot)$ extracts the real part.
Applying this function to Eq.\,\eqref{eq:received_signal_sum}, the BS can extract the sum of message vectors; when the estimation result for Eq.\,\eqref{eq:sum_true} is denoted as $\hat{\mathbf{r}}$, this operation can be written as,
\begin{align}
    \hat{\mathbf{r}} &= \mathrm{Re}\left(\frac{\mathbf{y}}{\sqrt{\rho}}\right) = \sum_{i=1}^{M} \mathbf{m}_i + \frac{\mathbf{z}_\mathrm{R}}{\sqrt{\rho}}\\
    &\triangleq \aircomp{\{\mathbf{m}_i\mid i=1, 2,\cdots, M\}},
    \label{eq:decoder-perfect}
\end{align}
where $\mathbf{z}_\mathrm{R} \triangleq \mathrm{Re}(\mathbf{z})$ and $\aircomp{\cdot}$ is the function for the AirComp-based summation operation.
For $\rho\rightarrow \infty$, the computation result follows $\hat{\mathbf{r}} \rightarrow \sum_{i=1}^{M} \mathbf{m}_i$.
\section{Purely Weighted Averaging AirComp}
\label{sect:pure_weighted_averaging}
Before introducing the proposed method, this section presents the pure weighted averaging operation with AirComp.
Then, we show the noise enhancement problem in the weighted averaging operation, which motivates us to design the adaptive weighted averaging.
\par
If the sum of the weights is normalized to 1 in advance of the AirComp, the weighted averaging operation can be realized by the weighted {\it sum}, which is given by $\sum_{i=1}^{M}w_{il}s_{il}$.
In contrast, the weighted {\it averaging} as in Eq.\,\eqref{eq:weighted_averaging} is often required in the case where the sum of the weights is not normalized and not available at the BS.
This operation must compute (i) the sum of weighted values and (ii) the sum of weights.
To this end, these two messages are transmitted in separate time slots, and each sum is computed based on AirComp.
\par
For the sum of weighted values and the sum of weights, the $i$-th node sets the following two messages:
\begin{align}
    \mathbf{m}^{(0)}_i &= [w_{i1}s_{i1}, w_{i2}s_{i2}, \cdots, w_{iL}s_{iL}]^{\mathrm{T}},\\
    \mathbf{m}^{(1)}_i &= [w_{i1}, w_{i2}, \cdots, w_{iL}]^{\mathrm{T}},
\end{align}
After $\{\mathbf{m}^{(0)}_i\mid i=1, 2, \cdots, M\}$ and $\{\mathbf{m}^{(1)}_i\mid i=1, 2, \cdots, M\}$ are aggregated with AirComp over two time slots, the BS can obtain the following two vectors,
\begin{align}
    \hat{\mathbf{r}}^{(t)} = \aircomp{\{\mathbf{m}^{(t)}_i\mid i=1, 2, \cdots, M\}}\,\,\text{for } t\in\{0, 1\},
    \label{eq:sum_weight}
\end{align}
where $\rho^{(t)}$ is a scalar at $t$ based on Eq.\,\eqref{eq:powercontrol_perfect}, which is constant among nodes.
Defining $\hat{\phi}_l$ as the estimated result of $\phi_l$, the weighted averaged value can be computed by
\begin{align}
    \hat{\phi}_l &= \frac{1}{r^{(1)}_l}r^{(0)}_l\\
    &= \frac{1}{\sum_{i=1}^{M}w_{il} + \frac{z^{(1)}_{\mathrm{R}l}}{\sqrt{\rho^{(1)}}}} \left(\sum_{i=1}^{M} w_{il} s_{il} + \frac{z^{(0)}_{\mathrm{R}l}}{\sqrt{\rho^{(0)}}}\right), \label{eq:pure_weighting_with_awgn}
\end{align}
where $r^{(t)}_l$ is the $l$-th element in $\hat{\mathbf{r}}^{(t)}$ and $z^{(t)}_{\mathrm{R}l}$ is the $l$-th element in $\mathbf{z}^{(t)}_\mathrm{R}$.
This computation result follows $\hat{\phi}_l \rightarrow \phi_l$ when $\rho^{(0)}\rightarrow \infty$ and $\rho^{(1)}\rightarrow \infty$.
However, since it divides the weighted sum by the sum of weights, its computation accuracy is often degraded severely due to $z^{(1)}_{\mathrm{R}l}$, as explained in Sect.\,\ref{sect:intro}.
\section{Adaptively Weighted Averaging AirComp}
\label{sect:proposed}
In this section, we propose an adaptive weighting method to improve the noise tolerance in the weighted averaging operation.
\subsection{Main Idea}
\begin{figure}[t]
    \centering
    \includegraphics[width=1.0\linewidth]{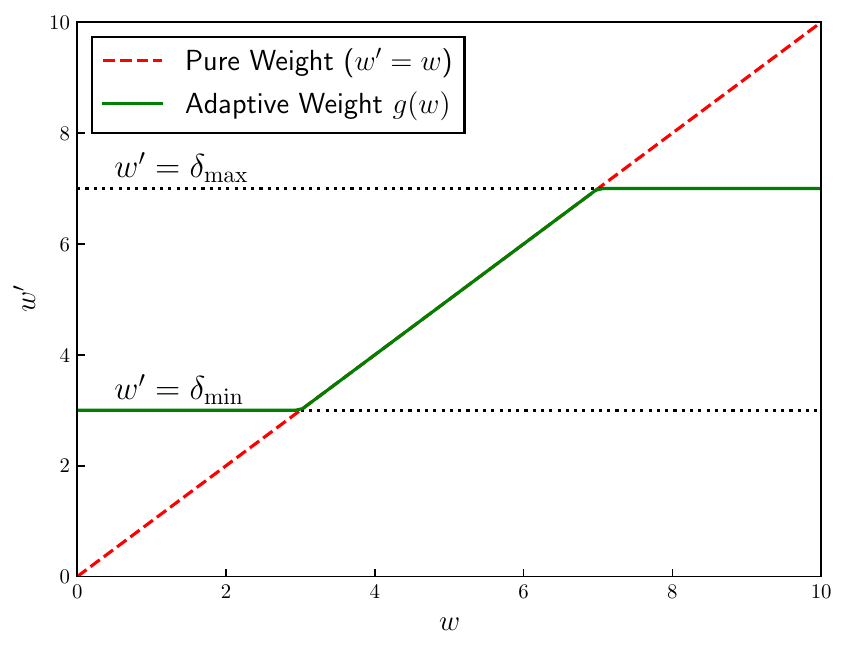}
    \caption{Behavior of adaptive weighting function ($\delta_\mathrm{max}=7.0$, $\delta_\mathrm{min}=3.0$).}
    \label{fig:example_adaptive_weight}
\end{figure}

\begin{figure}[t]
    \centering
    \includegraphics[width=1.0\linewidth]{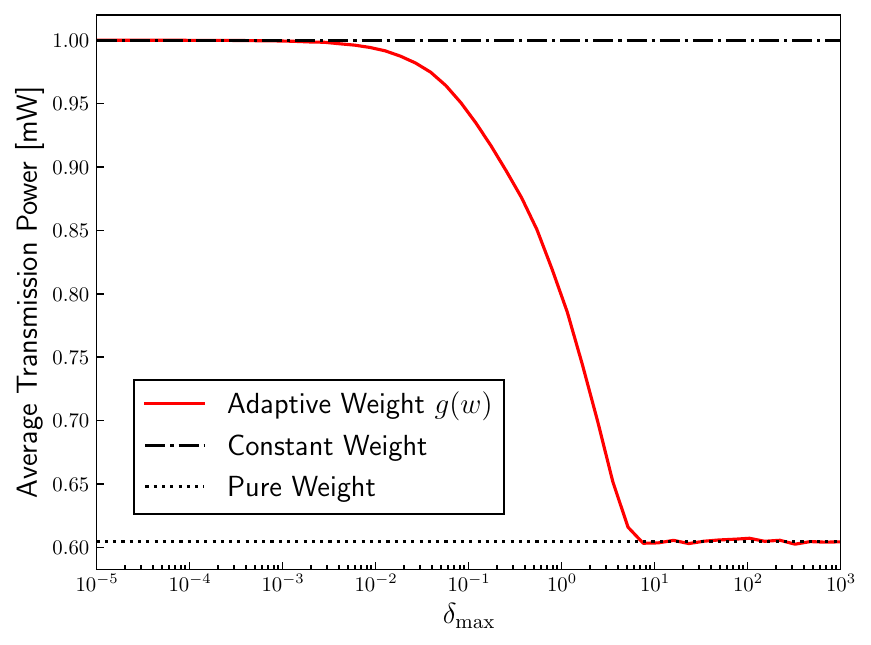}
    \caption{Effect of adaptive weighting on the average transmission power performance under AWGN channel in transmission step of $\mathbf{m}^{(0)}_i$($\delta_\mathrm{min}=0.0$, $P_\mathrm{max}=1.0$, $\overline{\gamma}_1 = \overline{\gamma}_2 = \cdots =1$ and $L=10$).
    Reducing $\delta_\mathrm{max}$ distorts $w$; however, it increases the average transmission power.}
    \label{fig:average_ptx_vs_delta_max}
\end{figure}
As can be seen from Eq.\,\eqref{eq:powercontrol_perfect}, in a pure weighted averaging operation, the power control parameter $\rho$ is designed so that the node having the minimum value of $\left(\sqrt{\overline{\gamma_i}}|h_i| / ||\mathbf{m}_i||\right)$ transmits the signal with the maximum transmission power $P_\mathrm{max}$.
When the channel coefficient $\sqrt{\overline{\gamma_i}}|h_i|$ is constant among the nodes, the average transmission power $\frac{1}{M}\sum_{i=1}^{M}||\mathbf{x}_i||^{2}$ decreases as the variance of $||\mathbf{m}_i||$ increases.
It often degrades the received power performance in the computation of Eq.\,\eqref{eq:sum_weight}, resulting in severe noise enhancement in computing Eq.\,\eqref{eq:pure_weighting_with_awgn}.
However, weighted averaging is crucial in various applications, such as environmental monitoring systems with sensing and spatial interpolation, where computation accuracy is essential.
\par
To mitigate this effect, our method truncates the weight vector $\mathbf{w}_i$ adaptively, based on the following function,
\begin{equation}
    g(w) = \max\left\{\min\left\{w, \delta_\mathrm{max}\right\}, \delta_\mathrm{min}\right\},
\end{equation}
where $\delta_\mathrm{max}$ and $\delta_\mathrm{min}$ are the upper and lower bounds of the weight, respectively.
Bounding $w$ with $\delta_{\max}$ and $\delta_{\min}$ reduces the variance of $\|\mathbf{m}_i\|$, resulting in less variation in transmission power. This, in turn, increases $\rho$, thereby enhancing computational accuracy.
\par
An example of this equation is illustrated in Fig.\,\ref{fig:example_adaptive_weight}. Controlling $\delta_\mathrm{max}$ and $\delta_\mathrm{min}$ can make the weighted averaging behavior.
For example, it can realize the pure weighted averaging as with Eq.\,\eqref{eq:pure_weighting_with_awgn} when $\delta_\mathrm{max}\rightarrow \infty$ and $\delta_\mathrm{min}=0$. In contrast, it nearly performs the simple averaging computation when $\delta_\mathrm{max}=\delta_\mathrm{min}$, which gains the average transmission power.
Fig.\,\ref{fig:average_ptx_vs_delta_max} demonstrates a numerical simulation result regarding the effects of the adaptive weighting on the average transmission power, $\frac{1}{M}\sum_{i=1}^{M}||\mathbf{x}_i||^{2}$, under the AWGN channel.
This figure assumes $f_s(x)\sim \mathcal{N}(0, 1)$ and $f_w(x) = \lambda \mathrm{exp}(-\lambda x) (x\geq 0)$ with $1/\lambda = 1$ for the demonstration purpose.
Reducing $\delta_\mathrm{max}$ to $\delta_\mathrm{min}$ makes the weights constant. The difference in $||\mathbf{m}_i||$ between the nodes is reduced, increasing the average transmit power; it will improve the noise enhancement in Eq.\,\eqref{eq:pure_weighting_with_awgn}.

\subsection{Detailed Protocol}
Alg.\,\ref{alg:adaptive} summarizes the detailed protocol of the proposed AirComp.
Its main protocol follows the pure weighted averaging in Sect.\,\ref{sect:pure_weighted_averaging}, except for the transformation of the weight coefficients by the function $g(w)$ and tuning parameters in $g(w)$.
The BS first adjusts $\delta_\mathrm{max}$ and $\delta_\mathrm{min}$ according to the average CSI $\{\overline{\gamma}_i \mid i=1, 2, \cdots, M\}$ so that the expected MSE can be minimized.
Since the objective function depends on $f_w$ and $f_s$, this problem cannot be derived in a closed-form expression; thus, we solve this problem based on BO, which will be described in Sect.\,\ref{subsect:bo}.
After the BS broadcasts these parameters to the nodes, the nodes convert the weighted sensing values and their weights as follows:
\begin{align}
    \mathbf{m}^{(0)}_{\mathrm{AW},i} &= [g(w_{i1})s_{i1}, g(w_{i2})s_{i2}, \cdots, g(w_{iL})s_{iL}]^{\mathrm{T}},\label{eq:proposed_m0}\\
    \mathbf{m}^{(1)}_{\mathrm{AW},i} &= [g(w_{i1}), g(w_{i2}), \cdots, g(w_{iL})]^{\mathrm{T}}\label{eq:proposed_m1}.
\end{align}
Then, the BS obtains the following vectors via AirComp:
\begin{align}
    \hat{\mathbf{r}}^{(t)}_\mathrm{AW} = \aircomp{\{\mathbf{m}^{(t)}_{\mathrm{AW},i}\mid i=1, 2, \cdots, M\}}\,\,\text{for } t\in \{0,1\}.
    \label{eq:sum_weight_adaptive}
\end{align}
Finally, BS computes the weighted averaging vector $\hat{\bm{\phi}}_\mathrm{AW}=\left[\hat{\phi}_{\mathrm{AW},1}, \cdots, \hat{\phi}_{\mathrm{AW},L} \right]^\mathrm{T}$, where its $l$-th element is given by
\begin{align}
    \hat{\phi}_{\mathrm{AW},l} &=
        \frac{1}{\hat{\mathbf{r}}^{(1)}_\mathrm{AW}}\hat{\mathbf{r}}^{(0)}_\mathrm{AW} \\
        &= \frac{1}{\sum_{i=1}^{M}g(w_{il}) + \frac{z^{(1)}_{\mathrm{R}l}}{\sqrt{\rho^{(1)}}}} \left(\sum_{i=1}^{M} g(w_{il}) s_{il} + \frac{z^{(0)}_{\mathrm{R}l}}{\sqrt{\rho^{(0)}}}\right).
    \label{eq:proposed_weighted_averaging}
\end{align}
Note that the messages $\mathbf{m}^{(t)}_{\mathrm{AW},i}$ converge to $\mathbf{m}^{(t)}_{i}$ (for $t \in \{0,1\}$) under the conditions $\delta_\mathrm{min}=0$ and $\delta_\mathrm{max}\to\infty$.
Further, the computation result $\hat{\phi}_{\mathrm{AW},l}$ converges to $\phi_l$ if $\rho^{(t)}\rightarrow \infty$, $\delta_\mathrm{min}=0$ and $\delta_\mathrm{max}\to\infty$.

\begin{algorithm}[t]
    \caption{Adaptively weighted averaging AirComp}
    \label{alg:adaptive}
    \begin{algorithmic}[1]
        \Require $f_w$, $f_s$, $\{\overline{\gamma}_i \mid i=1, 2, \cdots, M\}$, $\sigma_z^{2}$
        \State BS optimizes $\delta_\mathrm{max}$ and $\delta_\mathrm{min}$ via Alg.\,\ref{alg:bo}.
        \State BS broadcasts $\delta_\mathrm{max}$ and $\delta_\mathrm{min}$ to the nodes.
        \For{$t = 0 \text{ and } 1$}
            \For{$i = 1, 2, \cdots, M$}\Comment{For power control in Line 9.}
                \State The $i$-th node sets $\mathbf{m}^{(t)}_i$ by Eqs.\,\eqref{eq:proposed_m0}\eqref{eq:proposed_m1}.
                \State \parbox[t]{\dimexpr\linewidth-\algorithmicindent-8mm}{
                The $i$-th node sends $\|\mathbf{m}^{(t)}_i\|$ to the BS by digital transmission.
                }
                \State BS estimates $h^{(t)}_i$ when receiving $\|\mathbf{m}^{(t)}_i\|$.
            \EndFor
            \State BS controls $\rho^{(t)}$ and broadcasts it to the nodes.
            \State All nodes estimate $h^{(t)}_i$ when receiving $\rho^{(t)}$.
            \State All nodes construct $\mathbf{m}^{(t)}_{\mathrm{AW},i}$ by Eqs.\,\eqref{eq:proposed_m0}\eqref{eq:proposed_m1}.
            \State All nodes transmit $\mathbf{m}^{(t)}_i$ simultaneously.
            \State BS obtains $\hat{\mathbf{r}}^{(t)}_\mathrm{AW}$ by Eq.\,\eqref{eq:sum_weight_adaptive}.
        \EndFor

        \State The BS computes $\hat{\phi}_{\mathrm{AW},l}$ by Eq.\,\eqref{eq:proposed_weighted_averaging}.
        \Ensure $\hat{\bm{\phi}}_\mathrm{AW} = \bigl[\hat{\phi}_{\mathrm{AW},1}, \cdots, \hat{\phi}_{\mathrm{AW},L}\bigr]^\mathrm{T}$
    \end{algorithmic}
\end{algorithm}
 
\subsection{Optimizer for $\delta_\mathrm{max}$ and $\delta_\mathrm{min}$}
\label{subsect:bo}

\begin{algorithm}[t]
    \caption{Bayesian optimizer for \(\delta_\mathrm{max}\) and \(\delta_\mathrm{min}\)}
    \label{alg:bo}
    \begin{algorithmic}[1]
        \Require \(\mathcal{D}_\mathrm{BO}^{(0)} =\bigl\{\bigl[\theta^{(t)}, \hat{\epsilon}_\mathrm{AW}\bigl(\theta^{(t)}\bigr)\bigr]\mid t=1, 2, \cdots, N_\mathrm{init}\bigr\}\)

        \For{\(t = 1, 2, \cdots, T\)}
            \State \parbox[t]{\dimexpr\linewidth-\algorithmicindent-8mm}{
                Find \(\psi^{(t)}_\mathrm{opt}\) by solving Eq.\,\eqref{eq:problem-mle} via a gradient-based method.}
            \vspace{0.5em}
            \State Randomly sample a set of test inputs \(\Theta_\ast\).
            \State Calculate \(\mu(\theta_{\ast, i})\) by Eq.\,\eqref{eq:fullgpr-mean_bo} for \(\theta_{\ast, i} \in \Theta_\ast\).
            \State Calculate \(\sigma^2(\theta_{\ast, i})\) by Eq.\,\eqref{eq:fullgpr-var_bo} for \(\theta_{\ast, i} \in \Theta_\ast\).
            \State Calculate \(\alpha(\theta_{\ast, i})\) by Eq.\,\eqref{eq:expected_improvement} for \(\theta_{\ast, i} \in \Theta_\ast\).
            \State Find \(\theta^{(t+1)}\) by Eq.\,\eqref{eq:bo_nextinput}.
            \State Calculate \(\hat{\epsilon}_{\mathrm{AW}}\bigl(\theta^{(t+1)}\bigr)\).
            \State Update the dataset to \(\mathcal{D}_\mathrm{BO}^{(t+1)}\) by Eq.\,\eqref{eq:bo_nextdataset}
        \EndFor

        \Ensure \(\theta_\mathrm{opt} = \underset{\theta^{(t)} \in \Theta^{(T)}} {\operatorname{argmax}}\;\bigl(-\hat{\epsilon}_{\mathrm{AW}}\bigr)\bigl(\theta^{(t)}\bigr)\)
    \end{algorithmic}
\end{algorithm}

Alg.\,\ref{alg:adaptive} requires the optimization of $\delta_\mathrm{max}$ and $\delta_\mathrm{min}$ before the AirComp.
Assuming that $\theta = \{\delta_\mathrm{min}, \delta_\mathrm{max} \}$, this problem can be written as the MSE minimization in the adaptive weighted averaging, i.e.,
\begin{equation}
    \theta_\mathrm{opt} =
    \underset{\theta} {\operatorname{argmin}}\;\underbrace{\mathbb{E}\left[(\phi_l - \hat{\phi}_{\mathrm{AW},l})^2\right]}_{\triangleq \epsilon_{\mathrm{AW}} (\theta)},
    \label{eq:objective_msemin}
\end{equation}
where $\phi_l$ and $\hat{\phi}_{\mathrm{AW},l}$ are calculated by Eqs.\,\eqref{eq:weighted_averaging}\eqref{eq:proposed_weighted_averaging}, respectively.
However, deriving the exact expression for $\epsilon_\mathrm{AW}$ requires $f_s$ and $f_w$, although these functions depend on the sensing conditions.
For instance, in an environmental monitoring application with spatial interpolation, $f_s$ and $f_w$ depend on the spatial correlation of the sensing target and the distribution of node locations.
This makes $\epsilon_\mathrm{AW}$ a black-box function; thus, we need to treat this problem as an unconstrained black-box optimization, where typical methods such as gradient-based optimization are inefficient due to the absence of a closed-form solution.
\par
To this end, we introduce BO\cite{shahriari-ieee2016}, a black-box optimization framework for adaptive experimental design. BO models the set of observation values as a GP and interpolates the values at unobserved regions to identify better inputs. This method can be divided into initialization, function modeling by GPR, and parameter update based on the acquisition function.
We utilize BO because it efficiently optimizes the unconstrained black-box optimization problem with fast convergence and sample efficiency, enabling the BS to optimize $\delta_\mathrm{max}$ and $\delta_\mathrm{min}$ through offline simulation prior to AirComp.
In general, BO can rapidly optimize objective functions with low-dimensional input parameters.
The proposed truncation function has only two hyperparameters, thereby BO can efficiently find suitable parameter values.
Furthermore, BO exploits correlations among experimental data to estimate results in unexplored regions via a GP-based non-parametric regression.
As shown in Fig.\,\ref{fig:average_ptx_vs_delta_max}, these parameters continuously adjust the weighting behavior between simple averaging and pure weighted averaging, enabling BO to effectively model the relationship between the objective function and the input parameters.
\par
This algorithm is summarized in Alg.\,\ref{alg:bo}; we detail each step below.
Note that GPR is used within BO for modeling the objective function $\epsilon_\mathrm{AW}(\theta)$.
In contrast, the D-GPR introduced in Sect.\,\ref{sect:gpr} is employed for regression analysis of sensing results. It is important to recognize that these two applications of GPR address fundamentally different problems.

\subsubsection{Initialization}
In the beginning, this method performs several trials of simulations to construct an initial dataset
\begin{equation}
  \mathcal{D}_\mathrm{BO}^{(0)} =\left\{\left[\theta^{(t)}, \hat{\epsilon}_\mathrm{AW}\left(\theta^{(t)}\right)\right]\mid t=1, 2, \cdots, N_\mathrm{init}\right\},
  \label{eq:initial-dataset}
\end{equation}
where $N_\mathrm{init}$ is the number of initial trials and $\theta^{(t)}=\left\{\delta^{(t)}_\mathrm{max}, \delta^{(t)}_\mathrm{min}\right\}$ is the $t$-th initial input randomly selected from its feasible region. Furthermore,  $\hat{\epsilon}_\mathrm{AW}\left(\theta^{(t)}\right)$ is the estimated value for $\epsilon_\mathrm{AW}\left(\theta^{(t)}\right)$, which can be obtained by a Monte-Carlo integration with $f_s$, $f_w$, and $\theta^{(t)}$.

\subsubsection{Function Modeling by GPR}
This step models the distribution of $\hat{\epsilon}_\mathrm{AW}$ based on $\mathcal{D}_\mathrm{BO}^{(t)}$ and GPR.
GPR is a kernel-based non-parametric regression method that models the target function as a GP.
It first needs to tune its kernel structure from the training data to perform the function modeling accurately.
Let us denote the kernel function as $k_\mathrm{BO}\left(\theta^{(i)}, \theta^{(j)}\right)$, where $\theta^{(i)}=\left\{\delta^{(i)}_\mathrm{min}, \delta^{(i)}_\mathrm{max}\right\}$ and $\theta^{(j)} = \left\{\delta^{(j)}_\mathrm{min}, \delta^{(j)}_\mathrm{max}\right\}$. For example, if we utilize the radial basis function (RBF) with scaling and noise terms for the kernel, this function can be given by,
\begin{align}
    k_\mathrm{BO}\left(\theta^{(i)}, \theta^{(j)}\right) &= \beta_1 \exp\left(\beta_2 d^2\left(\theta^{(i)}, \theta^{(j)}\right)\right) + \beta_3,\\
    d(\theta^{(i)}, \theta^{(j)}) &\triangleq \sqrt{\left(\delta^{(i)}_\mathrm{min} - \delta^{(j)}_\mathrm{min}\right)^2 + \left(\delta^{(i)}_\mathrm{max} - \delta^{(j)}_\mathrm{max}\right)^2},
\end{align}
where $\beta_1, \beta_2, \beta_3$ are positive-valued hyper-parameters.
A set of the hyper-parameters, ${\bm \psi} = \left\{\beta_1, \beta_2, \beta_3\right\}$, can be tuned by maximizing the log-marginal likelihood function for multivariate normal distribution; i.e.,

\begin{equation}
    \psi^{(t)}_\mathrm{opt} = \underset{\psi} {\operatorname{argmax}}\,\,\left(\log p\left({\bm y}_\mathrm{BO}^{(t)}\mid \Theta^{(t)}, {\bm \psi}\right)\right),
    \label{eq:problem-mle}
\end{equation}
where
\begin{align}
  \log p\left({\bm y}_\mathrm{BO}^{(t)}\mid \Theta^{(t)}, {\bm \psi}\right) &=
  - \frac{1}{2}\log\mathrm{det}({\bm K}_\mathrm{BO}) -\frac{\left|\mathcal{D}^{(t)}_\mathrm{BO}\right|}{2}\log2\pi \nonumber \\
  -\frac{1}{2}\left({\bm y}_\mathrm{BO}^{(t)}-\overline{{\bm y}}_\mathrm{BO}^{(t)}\right)^\mathrm{T}&{\bm K}_\mathrm{BO}^{-1}\left({\bm y}_\mathrm{BO}^{(t)}-\overline{{\bm y}}_\mathrm{BO}^{(t)}\right),
  \label{eq:log-marginal}
\end{align}
and
\begin{align}
    {\bm y}_\mathrm{BO}^{(t)} &= \left[\hat{\epsilon}_\mathrm{AW}^{(1)}, \hat{\epsilon}_\mathrm{AW}^{(2)}, \cdots, \hat{\epsilon}_\mathrm{AW}^{(t+N_\mathrm{init})}\right]^\mathrm{T},\\
    \Theta^{(t)} &= \left\{\theta^{(i)} \mid i=1, 2, \cdots, t+N_\mathrm{init}\right\}.
\end{align}
Further, ${\bm K}_\mathrm{BO}\in \mathbb{R}^{\left|\mathcal{D}^{(t)}_\mathrm{BO}\right| \times \left|\mathcal{D}^{(t)}_\mathrm{BO}\right|}$ is the kernel matrix, where its element is\footnote{Typically, the kernel matrix is expressed as the sum of a noiseless kernel matrix ${\bm K}_\mathrm{BO}'$ and a matrix that accounts for the noise term $\sigma_{\epsilon}^2 {\bm I}$ (${\bm I}$ is $\mathcal{D}^{(t)}\times \mathcal{D}^{(t)}$ identity matrix and $\sigma_{\epsilon}^2$ is the noise variance). Following GPyTorch's implementation\cite{gardner2018gpytorch}, this paper integrates these to ${\bm K}_\mathrm{BO}\triangleq {\bm K}_\mathrm{BO}'+\sigma^2_\epsilon {I}$.} $k_\mathrm{BO}\left(\theta^{(i)}, \theta^{(j)}\right)$.
Finally, $\overline{{\bm y}}_\mathrm{BO}^{(t)}$ is a vector with $(t+N_\mathrm{init})$ elements.
Its $i$-th element $\overline{y}_\mathrm{BO}\left(\theta^{(i)}\right)$ represents the prior mean at $\theta^{(i)}$. Assuming that the prior mean is constant regardless of $\theta^{(i)}$, this mean can be calculated by
\begin{equation}
 \overline{y}_\mathrm{BO}\left(\theta^{(i)}\right)=\frac{1}{\left|\mathcal{D}_\mathrm{BO}^{(t)}\right|} \sum_{i=1}^{\left|\mathcal{D}_\mathrm{BO}^{(t)}\right|}\hat{\epsilon}_\mathrm{AW}^{(i)}, \forall i.
\end{equation}
\par
Maximizing Eq.\,\eqref{eq:log-marginal} can be realized by a gradient-based method such as limited-memory Broyden-Fletcher-Goldfarb-Shanno (L-BFGS) method\,\cite{L-BFGS}\footnote{Although a gradient-based method is widely used for this optimization, its result often falls into local optima since the log-likelihood function for the multivariate normal distribution is a multi-modal function\cite{gp-books-rasmussenw06}. An option for improving the optimization performance is to introduce a multi-start method, which repeats an optimization algorithm with different initial start points\cite{Marti2018}. However, the BO independently performs the MLE at each iteration to model the covariance performance in the latest dataset $\mathcal{D}_\mathrm{BO}^{(t)}$. This approach independently performs one-shot optimization at each iteration, suggesting that the local optima in each MLE do not overly affect the final optimization result.
To accumulate sufficient experiments within a short execution time, we design our approach based on the one-shot gradient-based method.}.
After this maximum likelihood estimation (MLE), the GPR predicts the distribution of the output at the unknown inputs 
\begin{equation}
 \Theta_\ast = \{\theta_{\ast,i}\mid i=1, 2, \cdots,N_\mathrm{test}\},
\end{equation}
 where $\theta_{\ast,i}$ is the $i$-th randomly-selected test point, and $N_\mathrm{test}$ is the number of test inputs to be interpolated.
For an unknown input $\theta_{\ast,i}$, its output mean and variance can be given by the following equations, respectively:
\begin{align}
  \mu (\theta_{\ast,i}) &=  \overline{y}(\theta_{\ast,i}) + {\bm k}^{\mathrm{T}}_{\ast,i}{\bm K}_\mathrm{BO}^{-1}\left({\bm y}^{(t)}-\overline{{\bm y}}^{(t)}\right) \label{eq:fullgpr-mean_bo}\\
  \sigma^2 (\theta_{\ast,i}) &= k_\mathrm{BO}\left(\theta_{\ast,i}, \theta_{\ast,i} \right) - {\bm k}^{\mathrm{T}}_{\ast,i}{\bm K}_\mathrm{BO}^{-1} {\bm k}_{\ast,i}, \label{eq:fullgpr-var_bo}
\end{align}
where
\begin{align}
  {\bm k}_{\ast,i} &= k_\mathrm{BO}\left(\Theta^{(t)}, \theta_{\ast,i}\right) \\
  &\triangleq \left[k_\mathrm{BO}(\theta^{(1)}, \theta_{\ast,i}), \cdots, k_\mathrm{BO}(\theta^{(t+N_\mathrm{init})}, \theta_{\ast,i}) \right]^{\mathrm{T}}.
\end{align}
Finally, the MSE at unknown input is modeled as
\begin{equation}
    \hat{\epsilon}_\mathrm{AW}\left(\theta_{\ast,i}\right) \sim \mathcal{N}\left(\mu \left(\theta_{\ast,i}\right), \sigma^2 \left(\theta_{\ast,i}\right)\right).
\end{equation}
We compute $\hat{\epsilon}_\mathrm{AW}\left(\theta_{\ast,i}\right)$ for all $\theta_{\ast,i} \in \Theta_\ast$.
\subsubsection{Parameter Update Based on Acquisition Function}
Next, we estimate the acquisition function, a criterion for evaluating the goodness of input parameters.
We employed expected improvement (EI), which is a popular acquisition function for BO defined as 
\begin{equation}
    \alpha(\theta_{\ast,i}) = \mathbb{E}\left[\max\left(( - \hat{\epsilon}_\mathrm{AW}(\theta_{\ast,i}))-( - \hat{\epsilon}_\mathrm{AW})^{+},0\right)\right],\label{eq:expected_improvement}
\end{equation}
where $( - \hat{\epsilon}_\mathrm{AW})^{+}$ is the maximum value of $- \hat{\epsilon}_\mathrm{AW}(\theta_{\ast,i})$ in $\mathcal{D}^{(t)}$.
After $\alpha(\theta_{\ast,i})$ is calculated for all $\theta_{\ast,i} \in \Theta_\ast$, we then determine the optimized input at the step $t$, i.e., the next input, as
\begin{equation}
    \theta^{(t+1)} = \underset{\theta_{\ast,i}\in \Theta_\ast} {\operatorname{argmax}}\;\alpha(\theta_{\ast,i}).
    \label{eq:bo_nextinput}
\end{equation}
After the MSE for this input $\hat{\epsilon}_\mathrm{AW}\left(\theta^{(t+1)}\right)$ is evaluated, the dataset for BO is updated as,
\begin{equation}
    \mathcal{D}_\mathrm{BO}^{(t+1)} \leftarrow \mathcal{D}_\mathrm{BO}^{(t)}\bigcup \left\{ \left[\theta^{(t+1)}, \hat{\epsilon}_{\mathrm{AW}}\left(\theta^{(t+1)}\right)\right]\right\}.
    \label{eq:bo_nextdataset}
\end{equation}
Steps 2 and 3 are iterated $T$ times; finally, the optimized input $\theta_\mathrm{opt}$ can be determined as the input point resulting in the minimum MSE, i.e.,
\begin{equation}
    \theta_\mathrm{opt} = \underset{\theta^{(i)} \in \Theta^{(T)}} {\operatorname{argmax}}\; \left(- \hat{\epsilon}_\mathrm{AW}\left(\theta^{(i)}\right)\right).
\end{equation}

\subsection{Communication Delay Performance}
Table\,\ref{table:required_timeslots} summarizes the time slots for the proposed and baseline methods at $L=1$.
Pure weighted averaging and adaptive weighted averaging require two time slots for AirComp to compute the weighted sum and the sum of weights, while simple averaging requires only one slot.
Further, the digital transmission requires $\mathcal{O}(2M)$ slots since it needs to collect $\mathbf{m}^{(0)}_i$ and $\mathbf{m}^{(1)}_i$ at the BS, and then the BS processes the received messages.
All AirComp-based methods can efficiently compute the weighted averaging compared to the digital transmission.

\subsection{Computational Complexity Performance}
BO requires training the kernel function, modeling the black-box function, and computing the acquisition function, resulting in increased computational complexity compared with non-adaptive methods.  
We discuss the additional complexity of the proposed method as follows.  
\par
To tune the kernel function, BO must maximize the log-marginal likelihood function in Eq.\,\eqref{eq:log-marginal}, which involves inverting an $(N_\mathrm{init}+t)\times (N_\mathrm{init}+t)$ matrix.  
Thus, when maximizing it using a gradient-based method, the computational complexity during the training phase is $\mathcal{O}(N_\mathrm{train}(N_\mathrm{init}+t)^3)$ at step $t$, where $N_\mathrm{train}$ represents the number of training iterations in the gradient-based method.
\par  
Subsequently, we need to predict the output mean and variance at the test points (Eqs.\,\eqref{eq:fullgpr-mean_bo}\eqref{eq:fullgpr-var_bo}) and compute the EI (Eq.\,\eqref{eq:expected_improvement}).
Since the inverse matrices in Eqs.\,\eqref{eq:fullgpr-mean_bo} and \eqref{eq:fullgpr-var_bo} can be derived from the kernel training results, computing these for a single input point requires $\mathcal{O}((N_\mathrm{init}+t)^2)$.  
If EI is calculated over $N_\mathrm{eval}$ test points, the computational complexity for prediction and EI calculation becomes $\mathcal{O}(N_\mathrm{eval}(N_\mathrm{init}+t)^2)$.
\par  
In summary, the computational complexity at step $t$, including both training and EI calculation, is $\mathcal{O}(N_\mathrm{train}(N_\mathrm{init}+t)^3 + N_\mathrm{eval}(N_\mathrm{init}+t)^2)$.  
Considering that the number of data points increases proportionally with $t$, the computational complexity for performing $T$ iterations of BO can asymptotically be expressed as:
\begin{equation}
 \mathcal{O}\left(N_\mathrm{train}T\left(N_{\mathrm{init}}+T\right)^3 + N_\mathrm{eval}T\left(N_{\mathrm{init}}+T\right)^2\right).
\end{equation}
\par
Note that BO needs to be performed only once before AirComp. Since the optimization results can be reused as long as the statistical CSI between the BS and nodes remains flat, the impact of this computational complexity on communication latency can be negligible in practice.

\begin{table}[t]
    \caption{Effects of $M$ on required time slots ($L=1$).}
    \label{table:required_timeslots}
    \centering
     \begin{tabular}{p{5.3cm}|p{2.7cm}}
      \hline
       Method & Required Time Slot \\\hline
       AirComp (pure weighted averaging) & $\mathcal{O}(1+1)$ \\
       AirComp (simple averaging) & $\mathcal{O}(1)$ \\
       AirComp (adaptive weighted averaging) & $\mathcal{O}(1+1)$ \\
       Digital transmission (pure weighted averaging) & $\mathcal{O}(2M)$ \\
      \hline
    \end{tabular}
\end{table}

\section{Application to Distributed GPR}
\label{sect:gpr}
\begin{figure*}[t]
    \centering
    \includegraphics[width=1.0\linewidth]{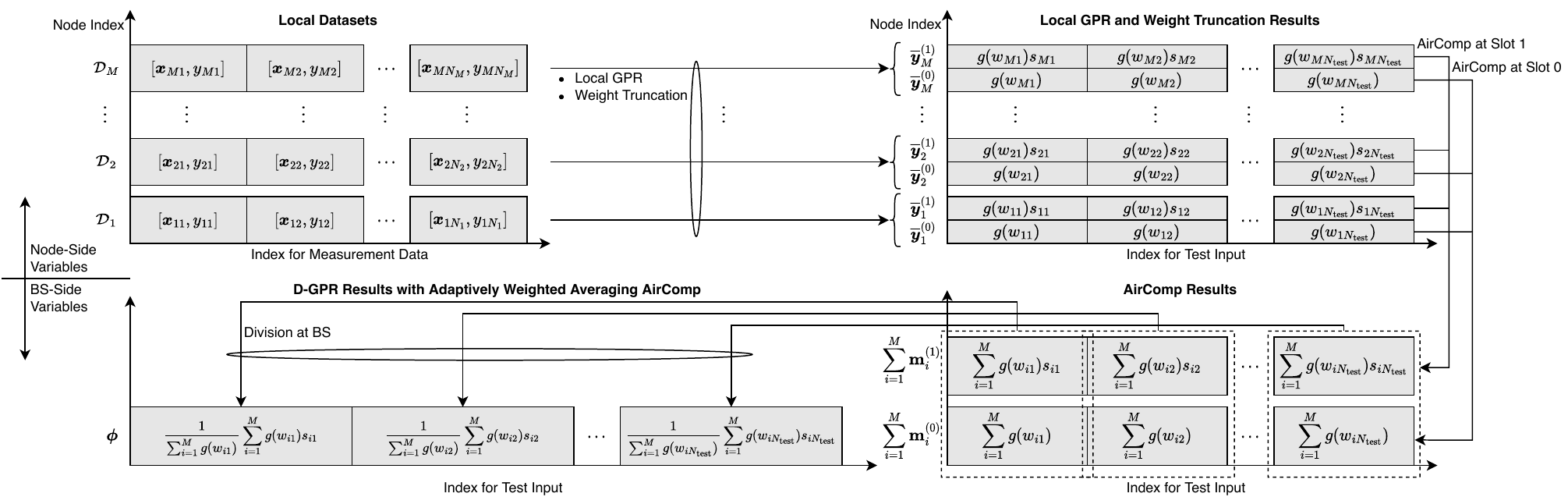}
    \caption{Relationship between variables in D-GPR with adaptively weighted averaging AirComp (noiseless case).}
    \label{fig:relationship_datasets_aircomp}
\end{figure*}
This section demonstrates how the proposed method can improve the actual computation problem; to this end, we apply the proposed method to a regression problem based on D-GPR.
There has been a wide range of applications of GPR, such as wireless sensor networks and radio map construction\cite{zeng-comst2024, bi-wirelesscommun2019}. Improving communication efficiency in the D-GPR will help such sensing applications.

\subsection{Task Definition}
Based on the definition in Sect.\,\ref{sect:systemmodel}, we consider a situation where $M$ sensing nodes are connected to a BS over wireless channels.
The $i$-th node has a dataset,
\begin{equation}
    \mathcal{D}_i = \left\{\left[{\bm x}_{in}, y_{in}\right]\mid n=1,2,\cdots,N_i\right\},
\end{equation}
where $N_i$ is the number of data, ${\bm x}_{in}$ is the input vector (e.g., sensing location), and $y_{in}=f({\bm x}_{in})+\epsilon$ is its output value generated from $\mathcal{N}(f({\bm x}_{in}), \sigma^2_\epsilon)$.
Further, $\epsilon \sim \mathcal{N}(0, \sigma^2_\epsilon)$ is the i.i.d. measurement noise; for instance, in a radio map construction task, ${\bm x}_{in}$ corresponds to the sensing location, $f({\bm x}_{in})$ represents the received signal power from the transmitter of interest, and $\epsilon$ accounts for noise caused by the sensor's temperature or signal interference from other transmitters.
\par
When local datasets are non-overlapped each other, the full dataset over the network can be expressed as
\begin{equation}
    \mathcal{D} = \bigcup_{i=1}^{M}\mathcal{D}_i,
\end{equation}
where the number of full data is $N=\sum_{i=1}^{M}N_i$.
Further, it is assumed that all data in $\mathcal{D}$ follows a GP: i.e.,
\begin{equation}
 f\sim \mathcal{GP}\left(\mu({\bm x}), k({\bm x}, {\bm x}')\right),
 \label{eq:gp}
\end{equation}
 where $\mathcal{GP}(\cdot, \cdot)$ is a PDF of GP, $\mu({\bm x})$ is the expectation value at ${\bm x}$, and $k({\bm x}, {\bm x}')$ is the kernel between ${\bm x}$ and ${\bm x}'$.
The task in this context is to estimate $f$ for test inputs 
\begin{equation}
 X_\ast = \{{\bm x}_{\ast,l} \mid l=1, 2, \cdots,N_\mathrm{test}\}
\end{equation}
 from $\mathcal{D}_i$ distributedly ($N_\mathrm{test}$ is equivalent to the message length $L$).
 Note that the relationship between $\mathcal{D}_i$, $\mathbf{m}_i^{(0)}$ and $\mathbf{m}_i^{(1)}$ is illustrated in Fig.\,\ref{fig:relationship_datasets_aircomp}.

\subsection{Preliminary of D-GPR Based on Product-of-Experts}
\label{subsec:pure-d-gpr}
We explain the estimation process on a test input ${\bm x}_{\ast,l}$ based on the full dataset $\mathcal{D}$ below; by applying the following process to all test inputs, regression analysis can be performed on $X_\ast$.
Consider a situation where BS has the full dataset $\mathcal{D}$ and performs the exact GPR.
 From the full dataset $\mathcal{D}$, we define ${\bm y}$ as a vector with $N$ elements that contains all $y_{in} \in \mathcal{D}$; i.e., ${\bm y}=[y_{11},y_{12},\cdots,y_{MN_M}]^{\mathrm{T}}$.
  The full GPR predicts the distribution of the output at the test input ${\bm x}_{\ast,l}$ as the Gaussian distribution with mean ($\mathbb{E}[f({\bm x}_{\ast,l})] = \mu ({\bm x}_{\ast,l})$) and variance ($\mathrm{Var}[f({\bm x}_{\ast,l})]= \sigma^2 ({\bm x}_{\ast,l})$) given by the following equations, respectively:
  \begin{align}
    \mu ({\bm x}_{\ast,l}) &=  \overline{y}({\bm x}_{\ast,l}) + {\bm k}^{\mathrm{T}}_{\ast,l}{\bm K} ^{-1}({\bm y}-\overline{\bm y}) \label{eq:fullgpr-mean}\\
    \sigma^2 ({\bm x}_{\ast,l}) &= k({\bm x}_{\ast,l}, {\bm x}_{\ast,l}) - {\bm k}^{\mathrm{T}}_{\ast,l}{\bm K}^{-1} {\bm k}_\ast, \label{eq:fullgpr-var}
  \end{align}
  where ${\bm K}\in \mathbb{R}^{N \times N}$ is the kernel matrix, where its element is $K_{ij} = k({\bm x}_i, {\bm x}_j)$ (${\bm x}_i$ is the $i$-th element in $X$).
  Further, ${\bm k}_\ast = k(X, {\bm x}_{\ast,l})$ and $\overline{\bm y}$ is a vector with $N$ elements, where its $i$-th element $\overline{y}({\bm x}_i)$ is the prior mean at ${\bm x}_i$; assuming that this mean is constant regardless of ${\bm x}_i$, this vector is given from $\overline{y}({\bm x}_1)=\overline{y}({\bm x}_2)=\cdots =\overline{y}({\bm x}_N)=\frac{1}{N}\sum_{i=1}^{N}y_i$, where $y_i$ is the $i$-th element in ${\bm y}$.
  \par
  However, performing the exact GPR at the BS requires the computational complexity following $\mathcal{O}(N^3)$ to calculate the $N\times N$ inverse matrix.
  D-GPR based on product-of-experts (PoEs), proposed in \cite{pmlr-v37-deisenroth15}, can improve this complexity issue by parallelizing computations of Eqs\,\eqref{eq:fullgpr-mean}\eqref{eq:fullgpr-var} to computations at distributed nodes.
  This method assumes that local dataset $\mathcal{D}_i$ is independent of each other\cite{pmlr-v37-deisenroth15}.
  Under this assumption, D-GPR estimates the statistical performance of $f({\bm x}_{\ast,l})$ based on the relationship,
 \begin{equation}
   p(f({\bm x}_{\ast,l})|\mathcal{D}) \approx \prod_{i=1}^M p(f({\bm x}_{\ast,l})|\mathcal{D}_i).
 \end{equation}
 To perform the D-GPR, each node estimates mean and variance based on the GPR for its local dataset; the $i$-th node computes the mean and variance on the $i$-th node, $\mu_i ({\bm x}_{\ast,l})$ and $\sigma^2({\bm x}_{\ast,l})$, based on the full GPR (Eqs.\,\eqref{eq:fullgpr-mean}\eqref{eq:fullgpr-var}) and its local dataset $\mathcal{D}_i$.
 Then, mean and variance for $p(f({\bm x}_{\ast,l})| \mathcal{D})$ can be calculated by the following equations, respectively.
 \begin{align}
    \mu^\mathrm{poe}({\bm x}_{\ast,l}) &= \left(\sigma^\mathrm{poe}({\bm x}_{\ast,l})\right)^2 \sum_{i=1}^{M}\left(\sigma_i({\bm x}_{\ast,l})\right)^{-2} \mu_i({\bm x}_{\ast,l}), \label{eq:poes-mean} \\
    \left(\sigma^\mathrm{poe}({\bm x}_{\ast,l})\right)^{-2}&= \sum_{i=1}^{M} \left(\sigma({\bm x}_{\ast,l})\right)^{-2}.\label{eq:poes-var}
 \end{align}
 
 This distributed processing requires each node to compute the inverse of $(N/M)\times (N/M)$ matrix. Thus, the computational complexity at the $i$-th node follows $\mathcal{O}\left((N/M)^3\right)$ when $N_1=N_2=\cdots=N_M$.
 However, with the digital transmission, its communication cost still depends on the number of nodes $M$ to aggregate the local GPR results $\mu({\bm x}_{\ast,l})$ and $\left(\sigma({\bm x}_{\ast,l})\right)^{-2}$.

\subsection{D-GPR with Adaptively Weighted Averaging AirComp}
As can be seen from Eqs.\,\eqref{eq:poes-mean}\eqref{eq:poes-var}, the D-GPR can be realized by the weighted averaging operation for $\mu_i({\bm x}_{\ast,l})$ and $\sigma^2_i({\bm x}_{\ast,l})$.
We can thus apply the AirComp-based weighted averaging to the D-GPR.
To exploit the adaptively weighted averaging, the nodes first prepare the messages $\mathbf{m}_i^{(0)}$ and $\mathbf{m}_i^{(1)}$ in Eqs.\,\eqref{eq:proposed_m0}\eqref{eq:proposed_m1} in which $w_{il}$ and $s_{il}$ can be written by
\begin{align}
    w_{il} = \left(\sigma_i ({\bm x}_{\ast, l})\right)^{-2},\,s_{il} = \mu_i\left({\bm x}_{\ast, l}\right).
\end{align}
Then, the BS can obtain the weighted averaging of $\mu_i({\bm x}_{\ast,l})$ and $\left(\sigma_i({\bm x}_{\ast,l})\right)^{-2}$ by Alg.\,\ref{alg:adaptive}.
\par
Note that Alg.\,\ref{alg:bo} in this method requires samples from PDFs regarding $w_{il}$ and $s_{il}$. However, these PDFs depend on the test input $X$.
To obtain $f_s$ and $f_w$, assuming that the PDF of $y_{in}$ is available, the BS emulates the D-GPR with offline computation in advance.
This process can be implemented in the following way:
\begin{enumerate}
    \item Initialize empirical datasets regarding $w_{il}$ and $s_{il}$ as $\mathcal{W}= \mathcal{S}=\emptyset$
    \item Generate pseudo dataset $\mathcal{D}'_i$ based on Eq.\,\eqref{eq:gp}.
    \item Perform local GPR for $\mathcal{D}'_i$ based on Eqs.\,\eqref{eq:fullgpr-mean}\eqref{eq:fullgpr-var}.
    \item Calculates pseudo data regarding $w_{il}$ and $s_{il}$ based on Eqs.\,\eqref{eq:poes-mean}\eqref{eq:poes-var}. Further, define them as $w'_{il}$ and $s'_{il}$, respectively.
    \item Append elements $w'_{il}$ and $s'_{il}$ into $\mathcal{W}$ and $\mathcal{S}$, respectively.
    \item Repeat the above steps a sufficient number of times.
\end{enumerate}
Finally, the BS can realize samples from $f_s$ and $f_w$ by randomly sampling elements from $\mathcal{W}$ and $\mathcal{S}$, respectively.
\subsection{Performance in Radio Map Construction Task}
\label{subsect:performance}
\subsubsection{Simulation Model}
\begin{table}[t]
    \caption{Simulation parameters.}
    \label{table:simulation_parameters}
    \centering
     \begin{tabular}{p{4.0cm}|p{4cm}}
      \hline
       Parameter & Detail \\\hline
       Kernel function in GPR & Radial basis function (RBF) with scaling and noise term\\
      $N_\mathrm{test} (=L)$ & 10\\
       $P_\mathrm{max}$ & 10\,[dBm]\\
       AWGN $\sigma^2_z$ & -90\,[dBm]\\
       $d_\mathrm{cor}$ & 100\,[m]\\
       $\sigma_\mathrm{dB}$ & 8.0\,[dB]\\
       $P_\mathrm{Tx}$ & 10\,[dBm]\\
       $\eta$ & 3.0\\
      \hline
     \end{tabular}
   \end{table}

\begin{figure*}[t]
    \centering
    \includegraphics[width=1.0\linewidth]{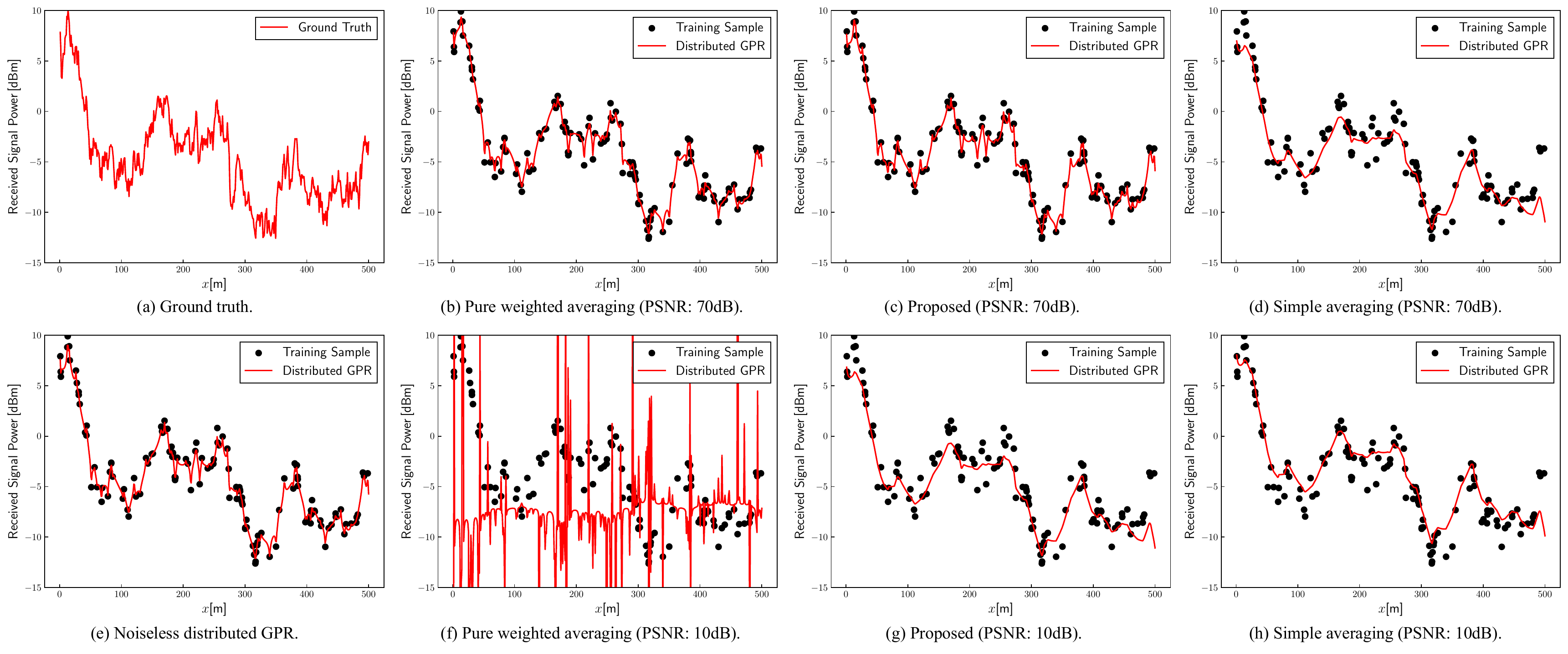}
    \caption{Examples of D-GPR in an 1D radio map construction task where $M=4$ and $N=128$ (AWGN channels).}
    \label{fig:example_dgpr}
\end{figure*}
We evaluate the performance of the AirComp-based D-GPR in a radio map construction task.
A radio map is a tool for visualizing the average received signal power values at an area of interest, which is essential for the design of wireless communication systems (e.g., resource allocation, location tracking, and spectrum sharing)\cite{bi-wirelesscommun2019, zeng-comst2024}.
Recent works have revealed that an accurate radio map can be constructed by sensing the received signal power values at multiple locations and spatial interpolation\cite{wang-jiot2020, sato-tvt2021}.
Since typical average received signal power values in outdoor scenarios can be modeled as a GP\cite{gudmundson_el1991}, GPR is a theoretically optimal regressor for this spatial interpolation problem.
\par
Our simulation considers a one-dimensional radio map construction task where the target transmitter (apart from the BS, which is the server role in AirComp) is located at $x_\mathrm{Tx} = -1$\,[m]; the measurement locations are sampled by $x_{i, l}\sim \mathcal{U}(1\,\text{[m]}, 500\,\text{[m]})$, and the BS estimates the received signal power values between $[1\,\text{[m]}, 500\,\text{[m]}]$.
\par
At the location $x$, the received signal power in the logarithmic form can be modeled as\cite{Goldsmith}
\begin{equation}
  P_\mathrm{Rx}(x) = P_\mathrm{Tx} - 10\eta \log_{10}||x_\mathrm{Tx}-x|| + \chi (x),
  \label{eq:radiomap_model}
\end{equation}
where $P_\mathrm{Tx}$ is the transmission power and $\eta$ is the path loss exponent.
Further, $\chi({\bm x})$ is the shadowing, which follows multivariate normal distribution with zero mean and standard deviation $\sigma_\mathrm{S}$\,[dB].
For two measurement locations $x_i$ and $x_j$, its spatial correlation can be modeled as an exponential function\cite{gudmundson_el1991}, i.e.,
\begin{equation}
  \mathrm{Cor}[\chi(x_i), \chi(x_j)] = \exp\left(-\frac{||x_i-x_j||}{d_\mathrm{cor}}\mathrm{ln}2\right),
\end{equation}
where $d_\mathrm{cor}$ is the correlation distance.
The task is to estimate the received signal powers at the unobserved locations by the D-GPR.
Other parameters are summarized in Table\,\ref{table:simulation_parameters}.

\subsubsection{Comparison Methods}
We compare the performance of the proposed AirComp with the several related methods, which are detailed below.

\begin{itemize}
    \item \textbf{Pure Weighted Averaging}: This method takes the weighted averaging based on Eq.\,\eqref{eq:pure_weighting_with_awgn}.
    \item \textbf{Simple Averaging}: In this method, the nodes transmit the non-weighted sensing values $\mathbf{m}^{(0)}_i = [s_{i1}, s_{i2}, \cdots, s_{iL}]$, and the BS aggregates them via AirComp. Then, the aggregated vector is divided by $M$ to obtain the averaged value. This operation can be given by,
    \begin{equation}
        \hat{\phi}_l = \frac{1}{M} \left(\sum_{i=1}^{M} s_{il} + \frac{z^{(0)}_{\mathrm{R}l}}{\sqrt{\rho^{(0)}}}\right).
        \label{eq:aircomp_simple_averaging}
    \end{equation}
    Unlike the proposed method, this method directly divides the AirComp result by the constant $M$ at the BS, completely eliminating any noise impact during the division process.
    \item \textbf{Computation-Optimal Policy (COP)}:
    This method is a theoretically optimal approach for the summation operation, as designed in \cite{liu_twc2020}.
    For a sum of messages $\sum^{M}_{i=1}{\bf m}_i$ and the corresponding decoding result, the BS optimizes the nomographic function to minimize the MSE for the summation operation; i.e.,
    \begin{mini!}
        {\mathrm{Dec}({\bf y}), \{\mathrm{Enc}({\bf m}_i)\}^{M}_{i=1}}
        {\mathbb{E}\left[\left(\sum^{M}_{i=1}{\bf m}_i - \mathrm{Dec}({\bf y})\right)^2\right]\label{eq:optimization-problem}}
        {\label{prb}} 
        {}
        \addConstraint{||{\bm x}_i||^2 \leq P_\mathrm{max},\;\forall i\label{subeq:constraint-power}}.
    \end{mini!}
    Since this method is designed for the summation operation, we applied it separately to AirComp for the sum of weights and AirComp for the sum of weighted sensing values.
    The results are then combined at the BS side.
    \item \textbf{Path Loss Estimation}: This method does not use any AirComp and averaging methods. Instead, it estimates the target as $\hat{\phi}_l = \mathbb{E}_{f_s}[x]$; in the radio map construction, this method estimates $P_\mathrm{Rx}(x)$ as
    \begin{equation}
     \hat{P}_\mathrm{Rx}(x)=P_\mathrm{Tx} - 10\eta \log_{10}||x_\mathrm{Tx}-x||.
    \end{equation}
    \item \textbf{Noiseless D-GPR}: This method performs pure D-GPR without any error/noise in the information aggregation, which expresses a limit of regression accuracy.
\end{itemize}

\begin{figure}[t]
  \centering
  \includegraphics[width=1.0\linewidth]{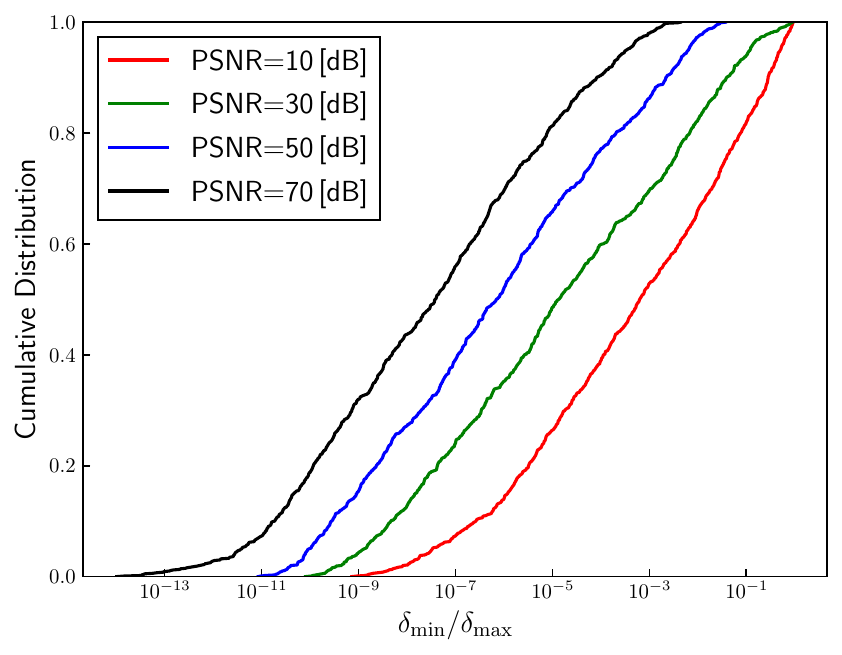}
  \caption{Effects of PSNR on the distribution of $\delta_\mathrm{min}/\delta_\mathrm{max}$ in Rayleigh fading channels where $M=32$ and $N=256$.\label{fig:cdf_deltamin_max}}
\end{figure}
\begin{figure*}[t]
  \centering
    \subfigure[AWGN, PSNR: 20\,dB.]{
      \includegraphics[width=0.3\linewidth]{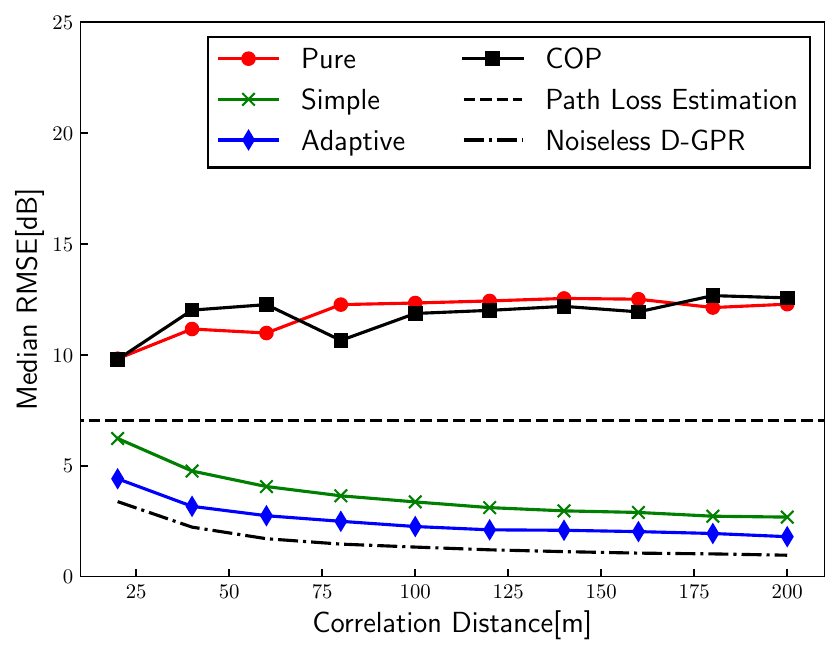}
    \label{subfig:rmse_vs_snr_awgn_snr20db}
    }
    \subfigure[AWGN, PSNR: 30\,dB.]{
      \includegraphics[width=0.3\linewidth]{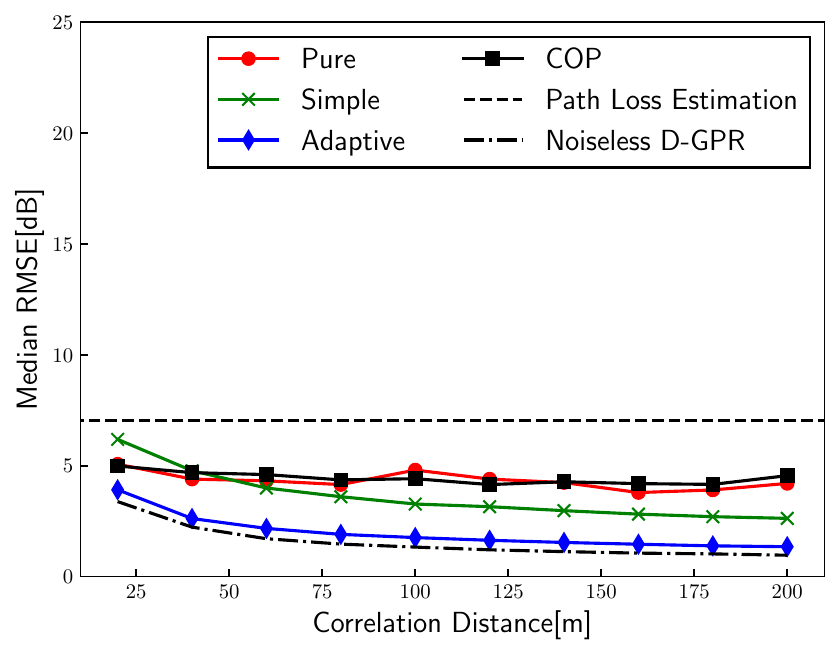}
    \label{subfig:rmse_vs_snr_awgn_snr30db}
    }
    \subfigure[AWGN, PSNR: 50\,dB.]{
      \includegraphics[width=0.3\linewidth]{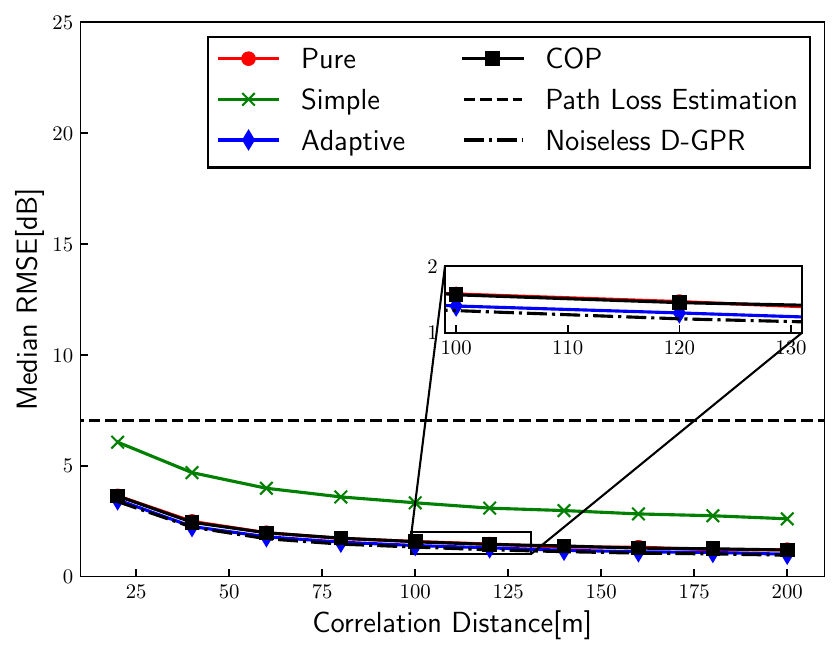}
      \label{subfig:rmse_vs_snr_awgn_snr50db}
    }
    \subfigure[Rayleigh fading, PSNR: 20\,dB.]{
      \includegraphics[width=0.3\linewidth]{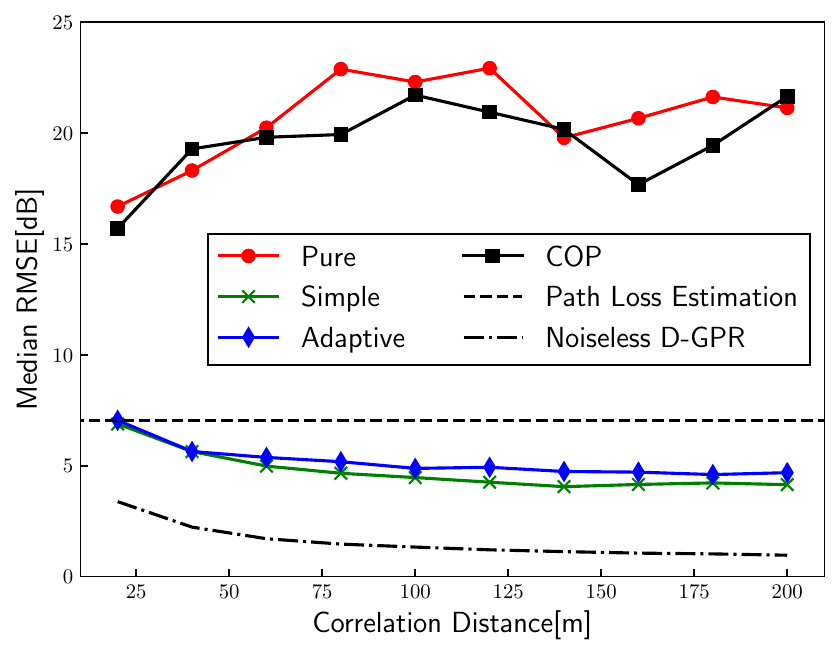}
      \label{subfig:rmse_vs_snr_fading_snr20db}
    }
    \subfigure[Rayleigh fading, PSNR: 30\,dB.]{
      \includegraphics[width=0.3\linewidth]{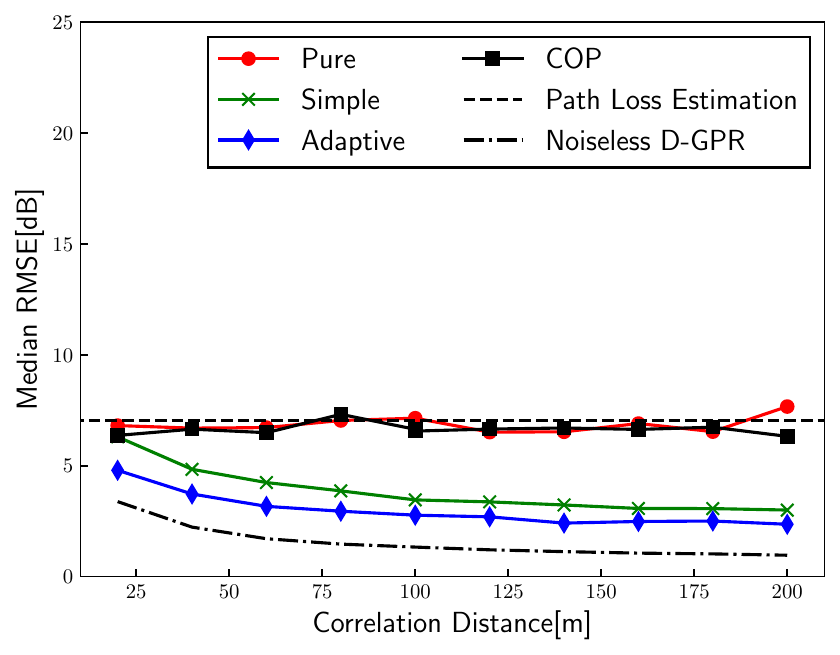}
      \label{subfig:rmse_vs_snr_fading_snr30db}
    }
    \subfigure[Rayleigh fading, PSNR: 50\,dB.]{
      \includegraphics[width=0.3\linewidth]{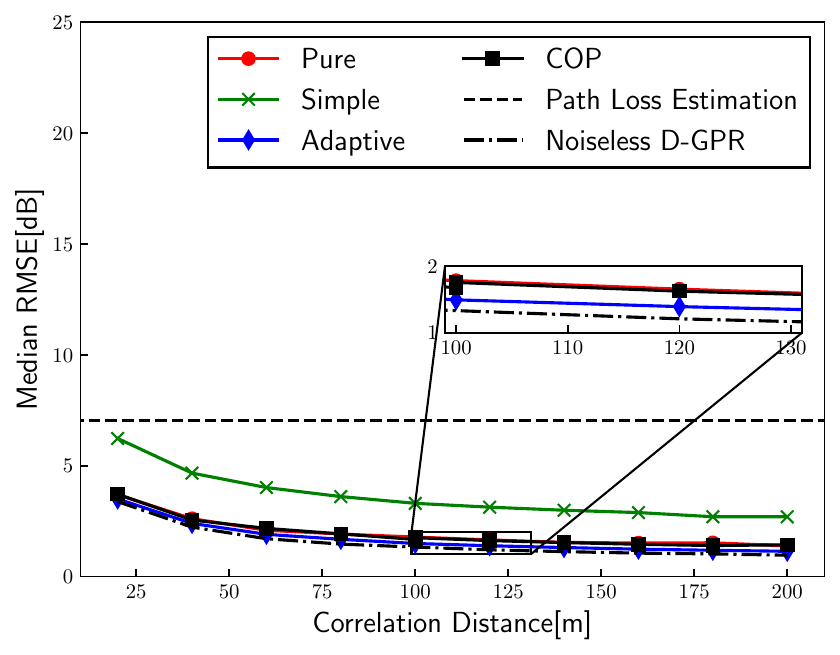}
      \label{subfig:rmse_vs_snr_fading_snr50db}
    }
  \caption{Effects of the correlation distance $d_\mathrm{cor}$ on RMSE performance where $M=32$.\label{fig:rmse_versus_dcor}}
\end{figure*}
\begin{figure*}[t]
  \centering
  \subfigure[AWGN, 8 nodes.]{
    \includegraphics[width=0.3\linewidth]{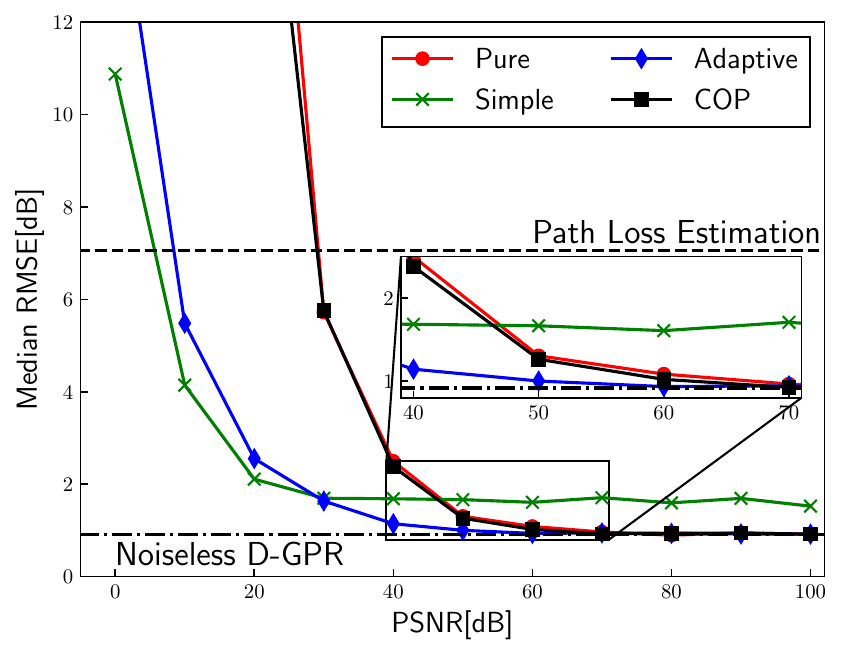}
  \label{subfig:8nodes_awgn}
  }
  \subfigure[AWGN, 16 nodes.]{
    \includegraphics[width=0.3\linewidth]{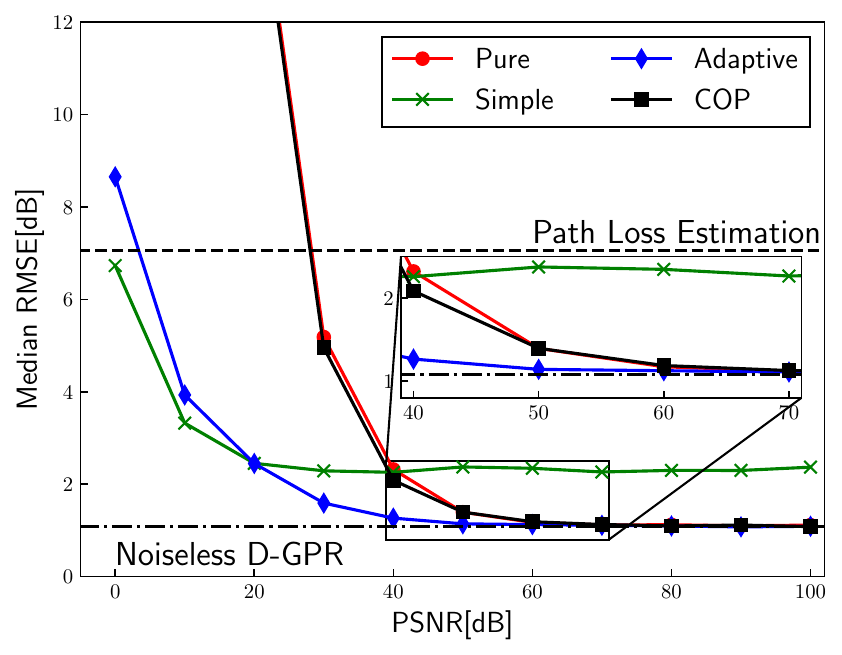}
    \label{subfig:16nodes_awgn}
  }
  \subfigure[AWGN, 32 nodes.]{
    \includegraphics[width=0.3\linewidth]{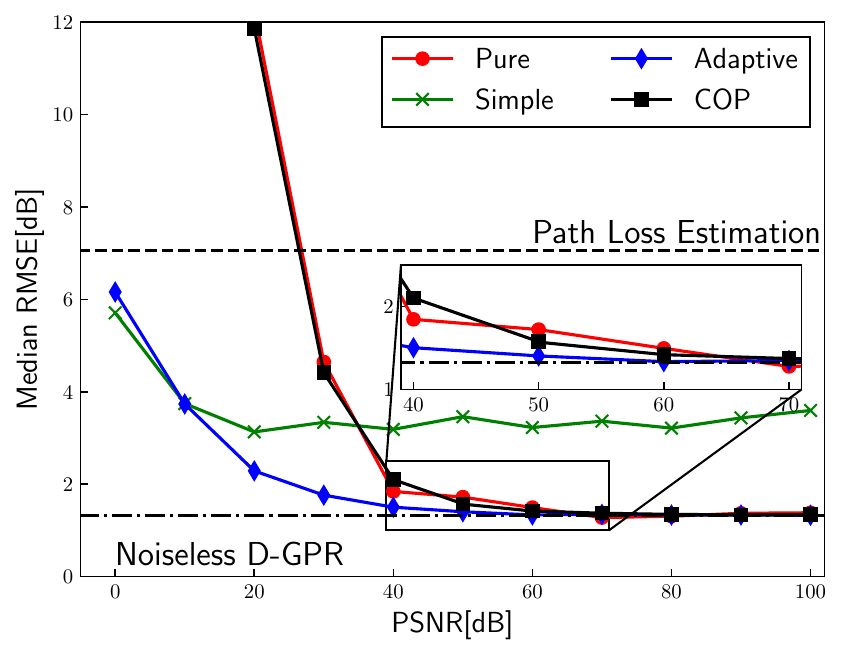}
    \label{subfig:32nodes_awgn}
  }
  \subfigure[Rayleigh fading, 8 nodes.]{
    \includegraphics[width=0.3\linewidth]{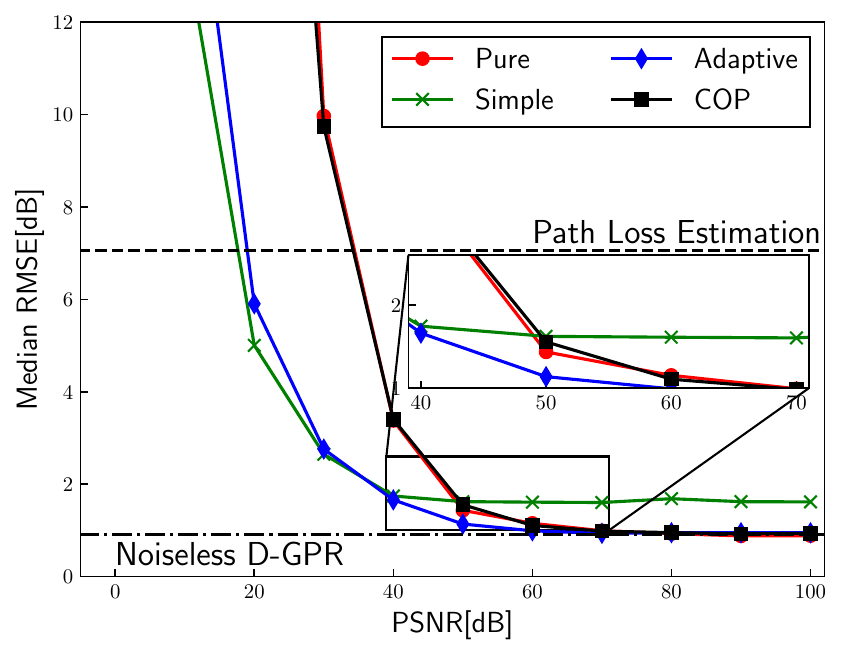}
    \label{subfig:8nodes_fading}
  }
  \subfigure[Rayleigh fading, 16 nodes.]{
    \includegraphics[width=0.3\linewidth]{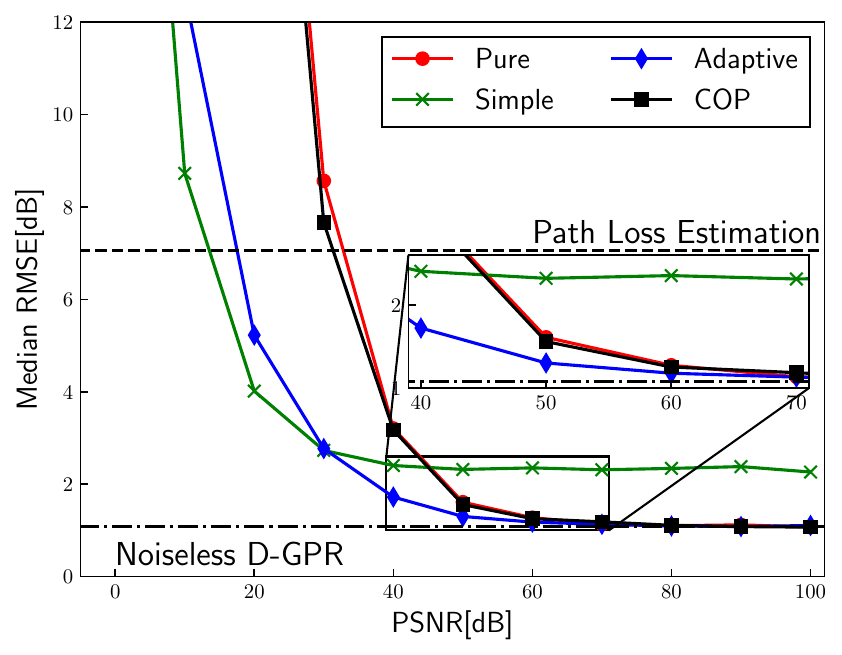}
    \label{subfig:16nodes_fading}
    }
  \subfigure[Rayleigh fading, 32 nodes.]{
    \includegraphics[width=0.3\linewidth]{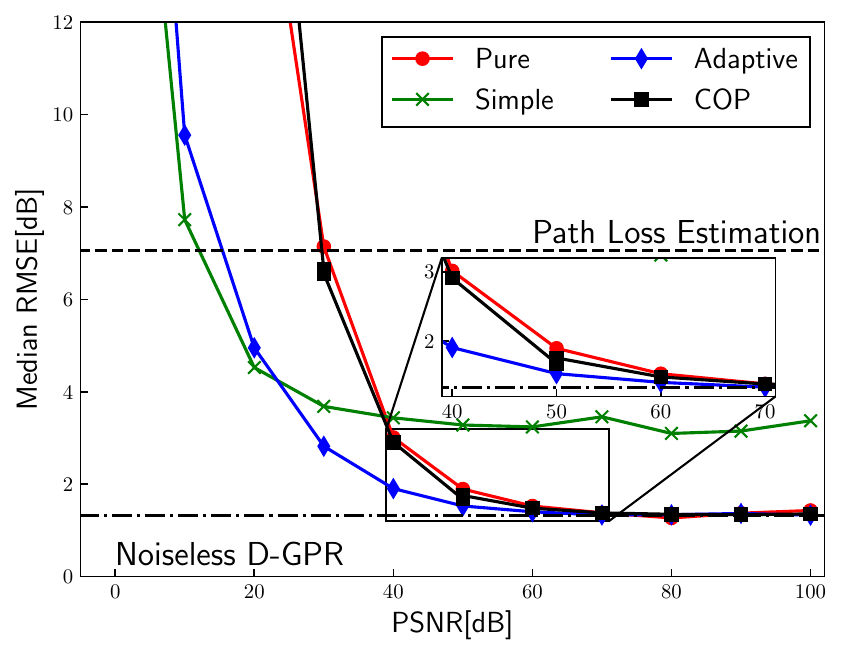}
  \label{subfig:32nodes_fading}
  }
\caption{RMSE versus PSNR where $N=256, d_\mathrm{cor}=100$.\label{fig:rmse-versus-snr}}
\end{figure*}
\subsubsection{Radio Map Construction Examples}
Effects of the averaging methods are exemplified in Fig.\,\ref{fig:example_dgpr}.
This figure demonstrates the radio map construction task in AWGN channels with peak signal-to-noise ratios (PSNRs) of 70\ dB and 10\ dB. $M=4$ nodes distributedly measure the received signal power at $N=128$ locations; i.e., each node has $N/M=32$ samples.
Note that, assuming $\overline{\gamma}_1 = \overline{\gamma}_2 = \cdots, =\overline{\gamma}$, this simulation defined the PSNR as the SNR with the maximum transmission power, i.e.,
\begin{equation}
    \mathrm{PSNR} \triangleq 10\log_{10}\left(\frac{P_\mathrm{max} \overline{\gamma}}{\sigma_z^2}\right)\,\text{[dB]}.
\end{equation}
\par
In $\mathrm{PSNR}=70\,\text{[dB]}$, pure weighted averaging and adaptive weighted averaging can accurately predict the ground truth as well as the noiseless D-GPR.
In contrust, although simple averaging can predict a rough relationship, it overly smooths the curve more than pure and adaptive weighted averaging methods, degrading the regression accuracy.
At $\mathrm{PSNR}=10\,\text{[dB]}$, pure weighted averaging has divergent predictions; in contrast, the results with adaptive weighted averaging and simple averaging are similar to the simple averaging at PSNR 70\,dB, which demonstrates the noise tolerant property.
\par
We next show the empirical cumulative distribution of $\delta_\mathrm{min}/\delta_\mathrm{max}$ in Fig.\,\ref{fig:cdf_deltamin_max}. We evaluated four PSNRs (70\,dB, 50\,dB, 30\,dB, and 10\,dB); each CDF was generated from 1000-times independent simulations.
As PSNR increased, the ratios tended to be lower: in terms of 90-percentile values, 
$\delta_\mathrm{min}/\delta_\mathrm{max} \approx 8.36 \times 10^{-5}$ at $\mathrm{PSNR}=70$, $2.22 \times 10^{-3}$ at $\mathrm{PSNR}=50$, $3.97 \times 10^{-2}$ at $\mathrm{PSNR}=30$ and $2.81 \times 10^{-1}$ at $\mathrm{PSNR}=10$, respectively.
This ratio approaches 1 when $\delta_\mathrm{min} \approx \delta_\mathrm{max}$, i.e., the truncation function converts the input to a constant weight, and AirComp result behaves as a pseudo simple averaging computation in a low-SNR situation. In contrast, this value approaches 0 when $\delta_\mathrm{min} \ll \delta_\mathrm{max}$. Thus, this method performs a pure weighted averaging computation in a high-SNR situation.

\subsubsection{Effects of $d_\mathrm{cor}$}
We discuss detailed RMSE performance.
This simulation assumed the number of test points was set as $N_\mathrm{test} = 10 (=L)$.
For a radio map construction simulation, the RMSE can be defined as
\begin{equation}
  \mathrm{RMSE} = \sqrt{\frac{1}{N_\mathrm{test}}\sum_{l=1}^{N_\mathrm{test}}\left(\phi_l - \hat{\phi}_l \right)^2}.
\end{equation}
After 10,000-times independent simulations, we obtained its median.
Note that, as exemplified in Fig.\,\ref{fig:example_dgpr}, the regression results with the pure weighted averaging method occasionally diverge due to the noise enhancement problem, leading to the divergence of the average RMSE. To discuss the statistical impact of PSNR, we use the median metric.
\par
Figs.\,\ref{subfig:rmse_vs_snr_awgn_snr20db}-\ref{subfig:rmse_vs_snr_awgn_snr50db} show effects of $d_\mathrm{cor}$ on the RMSE performance of D-GPR in AWGN where $N=256$.
The correlation between a test point and measurement samples increases as $d_\mathrm{cor}$ becomes long, thereby improving the RMSE in the noiseless D-GPR.
In the AWGN case at an PSNR of 50\,dB (Fig.\,\ref{subfig:rmse_vs_snr_awgn_snr50db}), both pure weighted averaging, computation-optimal method, and adaptive weighted averaging showed ideal RMSE equivalent to noiseless D-GPR in all correlation distances.
In contrast, simple averaging exhibited a degradation of approximately 1.6\,dB to 2.7\,dB compared to the noiseless D-GPR.
In the high PSNR region, RMSE can be improved by assigning a large weight to measurement samples near the test point.
Simple averaging always assigns equal weight to all samples, resulting in this gap.
\par
In contrast, at the PSNRs of 20\,dB and 30\,dB (Figs.\,\ref{subfig:rmse_vs_snr_awgn_snr20db} and \ref{subfig:rmse_vs_snr_awgn_snr30db}), simple averaging indicated higher accuracy than pure weighted averaging and COP for all correlation distances (Fig.\,\ref{subfig:rmse_vs_snr_awgn_snr20db}) and $d_\mathrm{cor}\geq 60\,\text{[m]}$ (Fig.\,\ref{subfig:rmse_vs_snr_awgn_snr30db}).
As shown in Eq.\,\eqref{eq:aircomp_simple_averaging}, this method divides the sum of non-weighted sensing values by $M$.
Since this operation avoids noise amplification, its RMSE was lower than the pure weighted averaging and COP at the PSNRs of 20\,dB and 30\,dB.
The adaptive weighted averaging also avoids noise amplification, as demonstrated in Fig.\,\ref{fig:example_dgpr}.
In the AWGN case, it further improved the RMSE by 0.8\,dB to 1.8\,dB at PSNR of 20\,dB and by 1.2\,dB to 2.2\,dB at PSNR of 30\,dB from the simple averaging method.
\par
Pure weighted averaging did not necessarily improve RMSE with increased correlation distance and showed lower accuracy than path loss estimation at the PSNR of 20 dB (e.g., 12.2\,dB at $d_\mathrm{cor}=200\,\text{[m]}$).
Further, COP demonstrated equivalent performance to the pure weighted averaging method under all simulation conditions.
Although this method minimizes the MSE for the sum of messages, it cannot mitigate the noise enhancement problem when computing the inverse of the sum, leading to inferior accuracy in low PSNR scenarios.
\par
Figs.\,\ref{subfig:rmse_vs_snr_fading_snr20db}-\ref{subfig:rmse_vs_snr_fading_snr50db} show effects of $d_\mathrm{cor}$ on the RMSE performance in Rayleigh fading channels where $N=256$.
As with the AWGN case, we evaluated the RMSEs at the PSNRs of 20\,dB (Fig.\ref{subfig:rmse_vs_snr_fading_snr20db}), 30\,dB (Fig.\ref{subfig:rmse_vs_snr_fading_snr30db}), and 50\,dB (\ref{subfig:rmse_vs_snr_fading_snr50db}).
The Rayleigh fading channel exhibited trends similar to those of the AWGN channel. 
In $\mathrm{PSNR} = 50\,\text{[dB]}$, pure weighted averaging, COP, and adaptive weighted averaging all achieved accuracy comparable to the noiseless case. 
However, as PSNR decreased, the accuracy of pure weighted averaging and COP degraded; e.g., at $\mathrm{PSNR} = 20\,\text{[dB]}$ and $d_\mathrm{cor}=200\,\text{[m]}$, the RMSE reached 21.1\,dB and 21.6\,dB, respectively.
In contrast, both simple averaging and adaptive weighted averaging demonstrated monotonic RMSE reduction with respect to $d_\mathrm{cor}$, regardless of PSNR. 
Unlike the AWGN case, at $\mathrm{PSNR} = 20\,\text{[dB]}$, adaptive weighted averaging and simple averaging showed comparable accuracy.
Nevertheless, throughout all conditions, adaptive weighted averaging consistently achieved the highest accuracy among AirComp-based methods, irrespective of PSNR or $d_\mathrm{cor}$.

\subsubsection{Effects of PSNR}
Next, Fig.\,\ref{fig:rmse-versus-snr} shows effects of PSNR on RMSE where $N=256, d_\mathrm{cor}=100\,\text{[m]}$.
This figure contains the results assuming $M=8, 16, 32$ in both AWGN and Rayleigh fading channels.
\par
In radio map construction task, comparing accuracy with the path loss estimation is essential to discuss how a method could estimate the shadowing fluctuation.
Focusing on pure weighted averaging and COP, it shows performance equivalent to noiseless D-GPR in the high PSNR region under any conditions. 
However, even at around $\mathrm{PSNR}=40\,\text{[dB]}$, a performance degradation of more than 1dB is observed.
For example, in Fig.\,\ref{subfig:8nodes_awgn}, the results are inferior to path loss estimation at PSNRs below 30\,dB.
On the other hand, both simple and adaptive weighted averaging methods are sufficiently advantageous compared to path loss estimation up to a PSNR of 20\,dB, achieving more than a 10\,dB improvement in PSNR compared to pure weighted averaging and COP.
  Note that in the low-SNR region, while adaptive weighted averaging approaches simple averaging, it is slightly inferior in terms of RMSE. 
  This is because simple averaging directly divides the AirComp result by the constant $M$ at the BS, completely eliminating any noise impact during the division process. 
  In contrast, adaptive weighted averaging divides the result by the sum of adaptive weights gathered through AirComp. 
  Although the sum of adaptive weights approaches $M$ at the low-SNR region, a slight residual noise impact remains during the division process due to computational errors in estimating $M$.
Furthermore, adaptive weighted averaging demonstrates excellent accuracy, asymptotically approaching that of noiseless D-GPR in the high PSNR region. 
Interestingly, the proposed method achieves higher accuracy than both simple averaging and pure weighted averaging in the mid-SNR region.
For instance, Fig.\,\ref{subfig:8nodes_awgn} shows that the proposed method outperforms the others in the PSNR range of 30 to 60\,dB, with gains of 0.54\,dB over simple averaging and 1.35\,dB over pure weighted averaging at a PSNR of 40\,dB.
This improvement is attributed to the use of weakly distorted weights, i.e., intermediate between fully distorted and ideal, which allows for increased average transmission power without significantly compromising the weight accuracy.
\par
This trend is consistent across both AWGN and Rayleigh fading channels and is independent of $M$.
With the proposed method, the RMSE converges to that of simple averaging in the low-SNR region, while remaining nearly equivalent to pure weighted averaging and COP in the high-SNR region.
Although a slight degradation in accuracy compared to simple averaging is observed in the low-SNR region, the proposed method maintains robust performance under all channel conditions.
Furthermore, the notable RMSE improvement in the mid-SNR region highlights the significant advantage of using a unified protocol to accommodate diverse AirComp scenarios.

\subsubsection{Effects of Unequal Average Channel Gain}
\label{subsec:unequal-avg-snr}
Figs.\,\ref{fig:rmse_versus_dcor} and \ref{fig:rmse-versus-snr} assumed $\gamma_1=\gamma_2=\cdots=\gamma$.
However, the average channel gain may not be equal between nodes in practice due to factors such as log-normal shadowing or differing distances from the transmitter.
Thus, we discuss the effects of unequal average channel gains on RMSE performance.
\par
This simulation extends the average channel gain of the $i$-th node as
\begin{equation}
  10\log_{10}\overline{\gamma}_i = 10\log_{10}\overline{\gamma} + \omega_i\,\text{[dB]},
\end{equation}
where $\omega_i$\,[dB] is the random variable following a normal distribution with zero mean and standard deviation $\sigma_\omega$\,[dB].
\par
\begin{figure}[t]
  \centering
    \subfigure[$10\log_{10}\overline{\gamma}_i = -70\,\text{[dB]}$.]{
      \includegraphics[width=0.45\linewidth]{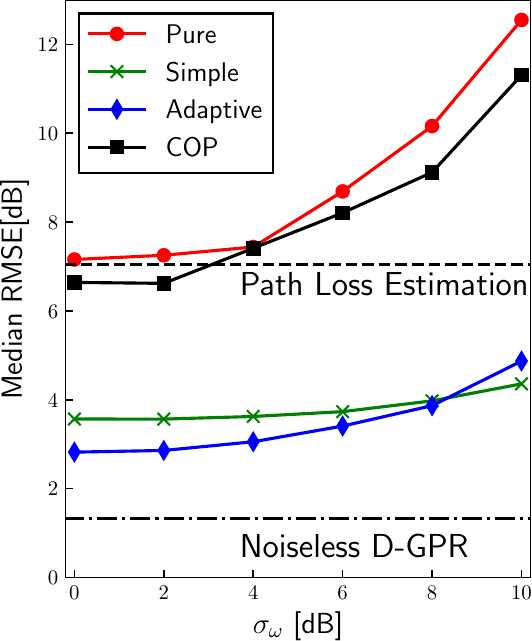}
      \label{subfig:effect_shadowing_70db}
    }
    \subfigure[$10\log_{10}\overline{\gamma}_i = -50\,\text{[dB]}$.]{
      \includegraphics[width=0.45\linewidth]{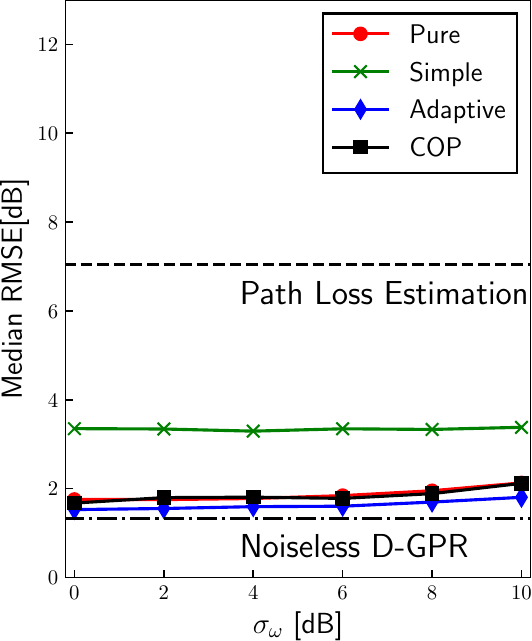}
      \label{subfig:effect_shadowing_50db}
    }
  \caption{Effects of unequal average channel gain on RMSE ($M=32, N=256$, Rayleigh fading).}\label{fig:effect_shadowing}
\end{figure}

Fig.\,\ref{fig:effect_shadowing} illustrates the effects of $\sigma_\omega$ on RMSE, where $M=32$ and $N=256$ in Rayleigh fading channels. 
We varied $\sigma_\omega$ from $0$\,[dB] to $10$\,[dB] to cover the typical range of shadowing standard deviation\,\cite{Goldsmith}. 
This evaluation selected $10\log_{10}\overline{\gamma}_i = -70\,\text{[dB]}$ as the condition under which adaptive weighted averaging exhibited significant accuracy improvement over pure weighted averaging in an equal average channel gain environment (Fig.\,\ref{subfig:effect_shadowing_70db}). 
In contrast, $10\log_{10}\overline{\gamma}_i = -50\,\text{[dB]}$ was chosen as the condition where both methods achieved accuracy comparable to noiseless D-GPR (Fig.\,\ref{subfig:effect_shadowing_50db}). 
It should be noted that the results for $\sigma_\omega = 0$ in Figs.\,\ref{subfig:effect_shadowing_70db} and \ref{subfig:effect_shadowing_50db} correspond to $\mathrm{PSNR} = 30$ and $\mathrm{PSNR} = 50$ in Fig.\,\ref{subfig:32nodes_fading}, respectively.
\par
In Figs.\,\ref{subfig:effect_shadowing_70db}, the RMSE of pure weighted averaging and COP degrades as $\sigma_\omega$ increases. 
For instance, at $\sigma_\omega = 10\,\text{[dB]}$, the RMSE values are 12.6\,[dB] and 11.3\,[dB], respectively. 
As shown by the PSNR dependency in Fig.\,\ref{subfig:32nodes_fading}, these methods experience significant RMSE degradation in $\mathrm{PSNR} < 50\,\text{[dB]}$, even in the equal channel gain case.
This degradation is due to an increase in low-SNR nodes as $\sigma_\omega$ increases.
Similarly, adaptive weighted averaging also shows degradation as $\sigma_\omega$ increases; however, it consistently outperforms pure weighted averaging and COP for all values of $\sigma_\omega$.
Although it performs slightly worse than simple averaging at $\sigma_\omega = 10$, the difference is less than 1\,dB, indicating that adaptive weighted averaging still provides sufficiently practical performance.

\subsubsection{Effects of Distribution Mismatching on RMSE}
\label{subsec:rmse-versus-mismatching}
Tuning of $\{\delta_\mathrm{min},\delta_\mathrm{max}\}$ with Alg.\,\ref{alg:bo} uses PDFs of sensing values ($f_s$) and their weights ($f_w$), requiring statistical information regarding $P_\mathrm{Rx}(x)$.
We discuss the effects of imperfect knowledge on the AirComp.
\par
Since the statistical performance of $P_\mathrm{Rx}(x)$ depends on the shadowing factor $\chi (x)$ in Eq.\,\eqref{eq:radiomap_model}, we model the imperfect knowledge on $P_\mathrm{Rx}(x)$ as the noisy correlation distance and standard deviation; i.e.,
\begin{align}
    d'_\mathrm{cor} &= d_\mathrm{cor} + \epsilon_d,\\
    \sigma'_\mathrm{dB} &= \sigma_\mathrm{dB} + \epsilon_\sigma,
\end{align}
where $\epsilon_d$ and $\epsilon_\sigma$ are the estimation error term for $d_\mathrm{cor}$ and $\sigma_\mathrm{dB}$, respectively.
Note that, to discuss the effects of imperfect optimization on the behavior of the proposed method, we assume that the BS performs BO with $d'_\mathrm{cor}$ and $\sigma'_\mathrm{dB}$, but D-GPR is computed based on $d_\mathrm{cor}$ and $\sigma_\mathrm{dB}$.
\par
\begin{figure}[t]
  \centering
  \subfigure[$\mathrm{PSNR}=30\,\text{[dB]}$]{
    \includegraphics[width=0.45\linewidth]{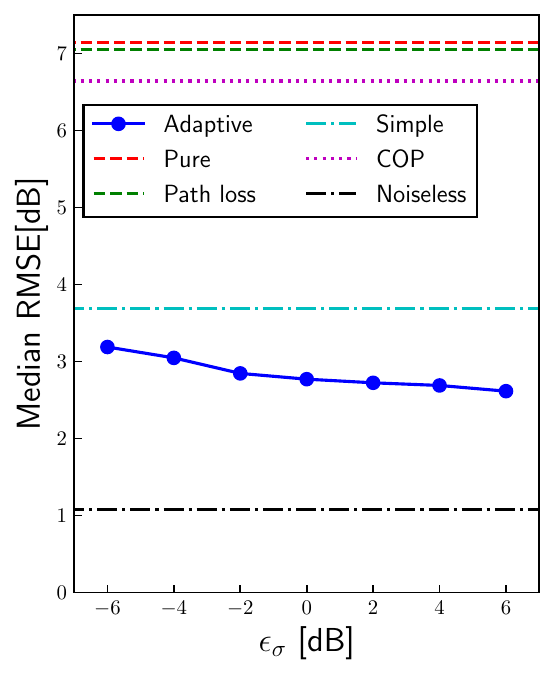}
    \label{subfig:effects_imperfect_stdev_30dB}
  }
  \subfigure[$\mathrm{PSNR}=50\,\text{[dB]}$]{
    \includegraphics[width=0.45\linewidth]{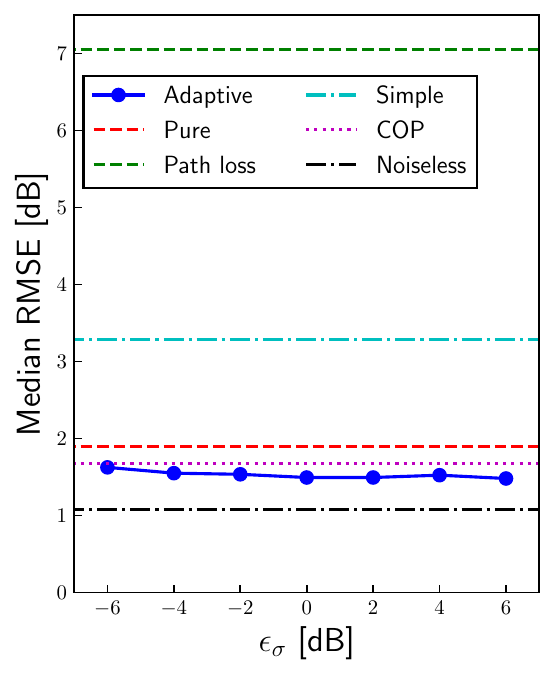}
    \label{subfig:effects_imperfect_stdev_50dB}
  }
\caption{Effects of distribution mismatching in standard deviation on RMSE in Rayleigh fading channels ($M=32, N=256, \epsilon_d = 0$).\label{fig:effects_imperfect_stdev}}
\end{figure}

Effects of $\epsilon_\sigma$ on RMSE is shown in Fig.\,\ref{fig:effects_imperfect_stdev}; $\mathrm{PSNR}=30\,\text{[dB]}$ (Fig.\,\ref{subfig:effects_imperfect_stdev_30dB}) and $\mathrm{PSNR}=50\,\text{[dB]}$ (Fig.\,\ref{subfig:effects_imperfect_stdev_50dB}).
We evaluated the performance under Rayleigh fading channels with parameters $M=32, N=256$, and $\epsilon_d=0$.
The performance at \(\epsilon_\sigma=0\) corresponds to that at $\mathrm{PSNR}=30$ and $50$ in Fig.\,\ref{subfig:32nodes_fading}.
At $\mathrm{PSNR}=30\,\text{[dB]}$, the RMSE of the proposed method degraded as the error factor $\epsilon_\sigma$ increased. 
However, the point where this method asymptotically approached the characteristics of noiseless D-GPR was consistent regardless of $\epsilon_\sigma$. 
For instance, even at $\epsilon_\sigma=-6.0$ (i.e., $\sigma'_\mathrm{dB}=2.0$), it achieved an accuracy improvement of 0.5\,dB compared to simple averaging. 
A similar trend was observed at $\mathrm{PSNR}=50\,\text{[dB]}$. 
Notably, in many wireless channels, $\sigma_\mathrm{dB}$ is approximately 12 dB\cite{Goldsmith}.
The simulation range for $\sigma'_\mathrm{dB}$ spans from 2 to 14\,dB, demonstrating that even when sufficient prior information on shadowing is unavailable, the hyperparameters can be empirically set, enabling BO to effectively adapt the weights in D-GPR.
\par
\begin{figure}[t]
  \centering
  \subfigure[$\mathrm{PSNR}=30\,\text{[dB]}$]{
    \includegraphics[width=0.45\linewidth]{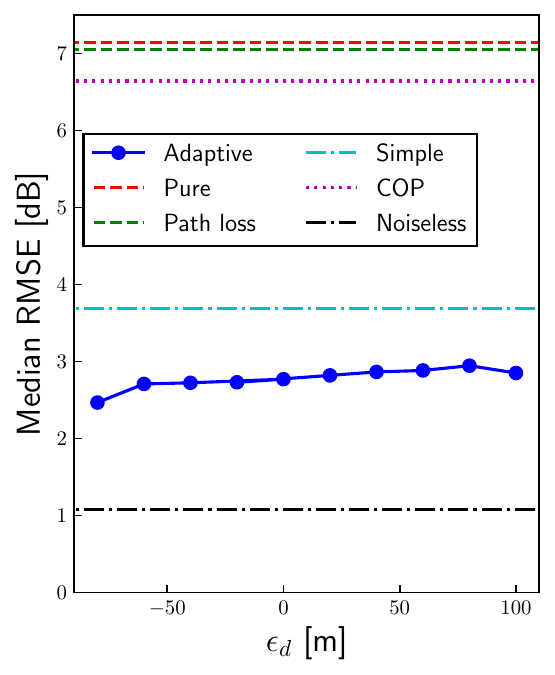}
  \label{subfig:effects_imperfect_dcor_30dB}
  }
  \subfigure[$\mathrm{PSNR}=50\,\text{[dB]}$]{
    \includegraphics[width=0.45\linewidth]{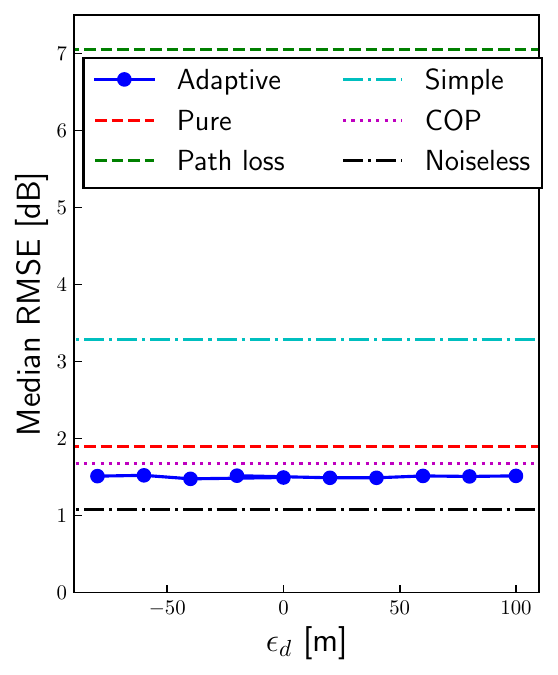}
    \label{subfig:effects_imperfect_dcor_50dB}
  }
\caption{Effects of distribution mismatching in spatial correlation distance on RMSE in Rayleigh fading channels ($M=32, N=256, \epsilon_\sigma = 0$).\label{fig:effects_imperfect_dcor}}
\end{figure}

Fig.\,\ref{fig:effects_imperfect_dcor} shows effects of $\epsilon_d$ on RMSE. As with Fig.\,\ref{fig:effects_imperfect_stdev}, we evaluated the performance in Rayleigh fading channels at $\mathrm{PSNR}=30\,\text{[dB]}$ (Fig.\,\ref{subfig:effects_imperfect_dcor_30dB}) and $\mathrm{PSNR}=50\,\text{[dB]}$ (Fig.\,\ref{subfig:effects_imperfect_dcor_50dB}); the results of compared GPR-based methods (simple averaging, pure weighted averaging, and noiseless D-GPR) indicate their performances with perfect knowledge on $d_\mathrm{cor}$ and $\sigma_\mathrm{dB}$.

At $\mathrm{PSNR}=30\,\text{[dB]}$, the RMSE of the adaptive weighted averaging method exhibited a gradual degradation as $\epsilon_d$ increased. 
However, similar to the influence of the estimation error in $\sigma_\mathrm{dB}$, the proposed method demonstrated the highest accuracy among AirComp-based approaches. 
Even at $\epsilon_d=100$, it showed an RMSE 0.8\,dB lower than that of simple averaging.
Overestimation of the correlation distance (i.e., $\epsilon_d > 0$) tends to assign higher weights to measurement samples far from the test point than the weights based on the accurate information on the correlation distance.
Nevertheless, the effect of noisy correlation distance only affects AirComp in terms of whether the AirComp behavior follows a purely weighted average or simple average.
The results show that the adaptive weighted averaging can work without significant accuracy degradation.
\par
These results suggest that the proposed method can realize accurate regression despite imperfect knowledge of $f_s$ and $f_w$.

\section{Case Study in FL Systems}
\label{sect:fl}
We demonstrated the effectiveness of the proposed AirComp-based weighted averaging through its application to D-GPR.
Given that weighted averaging is a fundamental operation in various domains beyond sensing, the proposed method has broad applicability.
\par
In FL systems, clients update their local model parameters based on private datasets and transmit them to the BS.
However, this process often incurs communication delays due to the transmission of high-dimensional model vectors from multiple clients.
While AirComp can mitigate these delays, using pure weighted averaging in AirComp degrades training performance.
Based on this background, this section evaluates the impact of the proposed method on training performance in AirComp-based FL systems.
\subsection{Federated Averaging with AirComp}
The task of FL is to train a machine learning model parameterized by a real-valued vector ${\bf v}$ based on iteration of training across distributed clients without disclosing their training dataset.
FedAvg, a typical algorithm for FL \cite{McMahan_AISTATS2017}, aims to construct a model that minimizes the global loss obtained by the weighted average of local loss values among clients.
Let us define $\mathcal{C}=\{1, 2, \cdots, K_\mathrm{KL}\}$ as the set of clients participating in the FL system.
When the $i$-th client has $N_{\mathrm{FL},i}$ training data, the objective is to find a model satisfying,
\begin{equation}
    {\bf v}^{\ast} \triangleq \underset{{\bf v}} {\operatorname{argmin}} \frac{1}{\sum^{K_\mathrm{FL}}_{i=1}N_{\mathrm{FL},i}}\sum_{i=1}^{K_\mathrm{FL}} N_{\mathrm{FL},i} F_i({\bf v}),
\end{equation}
where $F_i({\bf v})$ is the local loss at the $i$-th client.
For more detail, the FedAvg algorithm can be realized by the following process:
\begin{enumerate}
    \item BS selects $M_\mathrm{FL}(\leq K_\mathrm{FL})$ clients as the training clients at round $t_r$; we define the set of selected clients as $\mathcal{C}^{(t_r)}_\mathrm{S} \subseteq \mathcal{C}$.
    \item BS distributes a global model at round $t_r$, ${\bf v}^{(t_r)}$, to the selected clients.
    \item The selected clients distributedly update ${\bf v}^{(t_r)}$ to ${\bf v}_i^{(t_r)}$ based on their local datasets and a gradient descent-based optimization algorithm over several epochs.
    \item The selected clients report their updated models to the server.
    \item BS aggregates the local models based on the weighted averaging of the local models, i.e.\footnote{Depending on the implementation, model aggregation may be performed using a simple average instead of a weighted average. In such cases, the improvement effect of the proposed method cannot be expected.}, 
    \begin{equation}
        {\bf v}^{(t_r+1)} \leftarrow \frac{1}{\sum_{i \in \mathcal{C}^{(t_r)}_\mathrm{S}}N_{\mathrm{FL},i}} \sum_{i \in \mathcal{C}^{(t_r)}_\mathrm{S}} N_{\mathrm{FL},i} {\bf v}_i^{(t_r)}
        \label{eq:fl_aggregation}
    \end{equation}
    \item Server and clients iterate the above steps until the model training is sufficiently converged.
\end{enumerate}
Model aggregation in Eq.\,\eqref{eq:fl_aggregation} forms an averaging operation for the local models weighted by the number of the local training data.
Thus, we can apply Alg.\,\ref{alg:adaptive} for FedAvg by setting elements in Eqs.\,\eqref{eq:proposed_m0}\eqref{eq:proposed_m1} as,
\begin{align}
  w_{il} = N_{\mathrm{FL},i},\,s_{il} = v^{(t_r)}_{il},
\end{align}
where $v^{(t_r)}_{il}$ is the $l$-th element in ${\bf v}^{(t_r)}_i$.
AirComp reduces the number of slots required to aggregate local models to one, regardless of $M_\mathrm{FL}$, thereby improving communication efficiency in FedAvg.
Note that our implementation shrinks $\mathbf{m}^{(1)}_{\mathrm{AW},i}$ to one factor only since the weights in the $i$-th client follow $N_{\mathrm{FL},i}$ to avoid redundant information.

\begin{figure}[t]
  \centering
    \subfigure[$10\log_{10}\overline{\gamma}_i = 0\,\text{[dB]}$.]{
      \includegraphics[width=0.45\linewidth]{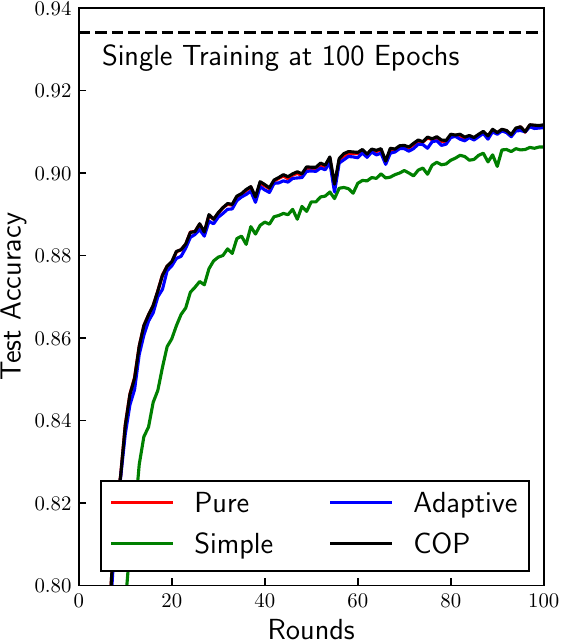}
      \label{subfig:fl_acc_0db}
    }
    \subfigure[$10\log_{10}\overline{\gamma}_i = -50\,\text{[dB]}$.]{
      \includegraphics[width=0.45\linewidth]{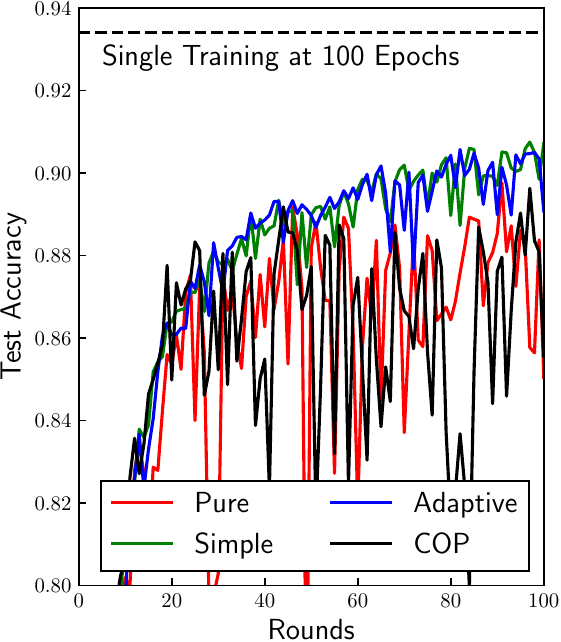}
      \label{subfig:fl_acc_50db}
    }
  \caption{Test accuracy of AirComp-based FL systems in an image classification task using the MNIST dataset.}\label{fig:acc_fl}
\end{figure}

\subsection{Image Classification Performance}
Based on a prior work in \cite{amiri-twc2020}, we conducted experiments on the MNIST dataset, which consists of 60,000 training samples and 10,000 test samples.
The images were trained using a single-layer feedforward neural network with ten neurons in the hidden layer (i.e., $|\mathbf{v}| = 7,960$).
The neurons were activated by the rectified linear unit function, and clients performed local training using stochastic gradient descent with one epoch per round.
In practice, the number of training samples per client is often imbalanced; for example, only certain clients may have access to large amounts of data.
To model such a scenario, we determined $N_{\mathrm{FL},i}$ based on random numbers generated from an exponential distribution with a mean of $60,000/K_\mathrm{FL}$, processed through the floor function.
Allowing for data overlap between clients, each client randomly sampled their data from the 60,000 training samples.
\par
Fig.\,\ref{fig:acc_fl} shows the test accuracy of AirComp-based FL systems in the image classification task under two conditions: a high-SNR environment (Fig.\,\ref{fig:acc_fl}(a)) and a noisy environment (Fig.\,\ref{fig:acc_fl}(b)) with AWGN channels. The test accuracy was evaluated per training round and averaged over ten independent simulations.
Additionally, the figure includes the following baseline strategies: (a) pure weighted averaging, and (b) simple averaging. The pure weighted averaging method and the COP perform AirComp without BO, while the simple averaging method computes the non-weighted sum of local models via AirComp and divides it by $M_\mathrm{FL}$.
\par
In Fig.\,\ref{subfig:fl_acc_0db}, the pure weighted averaging method and the COP outperform the simple averaging method throughout the training process. 
However, under the noisy condition shown in Fig.\,\ref{subfig:fl_acc_50db}, their accuracy values degrade significantly, and the simple averaging method outperforms it.
In contrast, the proposed method achieves performance equivalent to the pure weighted averaging method and the COP in the high-SNR case and the simple averaging method in the noisy case.
This case study demonstrates that the proposed adaptive weighting strategy behaves as a pseudo pure weighted averaging computation under high-SNR conditions, while acting as a pseudo simple averaging computation under low-SNR conditions, thereby enabling accurate distributed training across varying channel conditions.

\section{Conclusion}
\label{sect:conclusion}
This paper presented an adaptive weighting method for noise-tolerant weighted averaging in AirComp.
The proposed method truncates the maximum and minimum weights to balance noise tolerance and computation accuracy.
We developed an optimizer for the truncation function based on BO, which works in a black-box manner with the channel condition.
\par
The proposed method was applied to D-GPR for low-latency and accurate distributed regression.
The performance evaluation in a radio map construction task showed that the proposed method could improve the estimation accuracy of the D-GPR; its RMSE performance approaches the pure weighted average method in the high SNR region and the noise-tolerant simple average method in the low SNR region.
\par
Further, we presented a case study of the proposed method for FL systems. Similar to the D-GPR, it was demonstrated that the proposed method can adapt the model aggregation strategy based on the statistical channel conditions, thereby improving the training performance.
The proposed will enable accurate, low-latency weighted averaging in various applications over wireless networks.

\section*{Acknowledgement}
The authors would like to express their gratitude for the stimulating discussions during the meeting of the Cooperative Research Project at the Research Institute of Electrical Communication, Tohoku University.

\ifCLASSOPTIONcaptionsoff
  \newpage
\fi

\bibliographystyle{IEEEbib}
\bibliography{reference}

\begin{IEEEbiography}[{\includegraphics[width=1in,height=1.25in,clip,keepaspectratio]{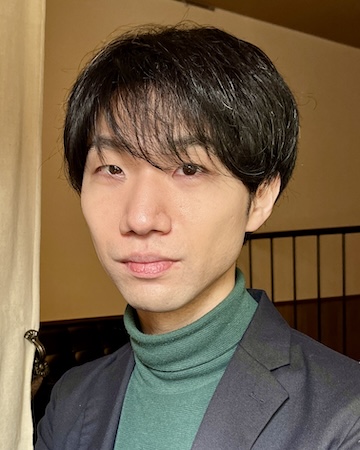}}]{Koya Sato}
  (Member, IEEE) received the B.E. degree in electrical engineering from Yamagata University, in 2013, and the M.E. and Ph.D. degrees from The University of Electro-Communications, in 2015 and 2018, respectively. From 2018 to 2021, he was an Assistant Professor with the Tokyo University of Science. He is currently an Assistant Professor with the Artificial Intelligence eXploration Research Center, The University of Electro-Communications. His current research interests include wireless communication, distributed machine learning, and spato-temporal statistics.
\end{IEEEbiography}

\begin{IEEEbiography}[{\includegraphics[width=1in,height=1.25in,clip,keepaspectratio]{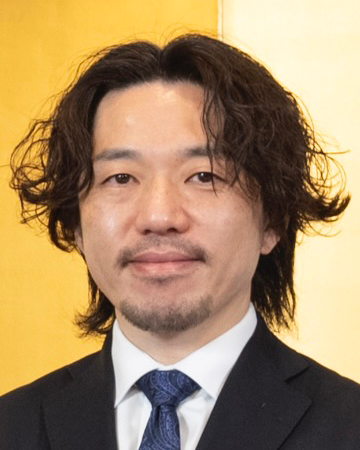}}]{Koji Ishibashi} (Senior Member, IEEE) received the B.E. and M.E. degrees in engineering from The University of Electro-Communications, Tokyo, Japan, in 2002 and 2004, respectively, and the Ph.D. degree in engineering from Yokohama National University, Yokohama, Japan, in 2007. 
From 2007 to 2012, he was an Assistant Professor at the Department of Electrical and Electronic Engineering, Shizuoka University, Hamamatsu, Japan. Since April 2012, he has been with the Advanced Wireless and Communication Research Center (AWCC), The University of Electro-Communications, Tokyo, Japan where he is currently a Professor. From 2010 to 2012, he was a Visiting Scholar at the School of Engineering and Applied Sciences, Harvard University, Cambridge, MA. He is a senior member of IEICE and IEEE. He is a recipient of Takayanagi Research Encouragement Award in 2009 and KDDI Foundation Award in 2023. He was certified as an Exemplary Reviewer of IEEE Communications Wireless Letters in 2015 and awarded by the Telecommunication Technology Committee (TTC) for his devotion to standardization activities in 2020. He has been an Associate Editor for IEEE Internet of Things Journal since 2025. He in the past served as an Associate Editor for IEICE Transactions on Communications, an Associate Editor for IEEE Journal on Selected Areas in Communications (JSAC) and a Guest Editor for IEEE Open Journal of the Communications Society. His current research interests are beamforming design, grant-free access, energy-harvesting, compressed sensing, coding, and MIMO technologies.
\end{IEEEbiography}

\end{document}